\documentclass[12pt, a4paper]{article}

\usepackage{subfigure}
\usepackage{array}
\usepackage{graphicx}
\usepackage{amsmath}
\usepackage{amssymb}
\usepackage{mathrsfs}
\usepackage{bm}
\usepackage{color}
\usepackage{slashed}
\usepackage[amsmath, thmmarks, hyperref]{ntheorem}
\usepackage{threeparttable}
\usepackage{appendix}
\usepackage[colorlinks, linkcolor=black, anchorcolor=black, citecolor=black]{hyperref}
\usepackage{soul}

\title{Weak Decays of $J/\psi$ and $\Upsilon(1S)$}

\author{Tianhong Wang$^a$\footnote{thwang@hit.edu.cn},~~Yue Jiang$^a$,~Han Yuan$^a$,~Kan Chai$^b$
\\
and~~Guo-Li Wang$^a$\footnote{gl\_wang@hit.edu.cn}\\
{\it \small   $^a$Department of Physics, Harbin Institute of Technology,
Harbin, 150001, China}
\\ \it\small  $^b$Applied Science College, Harbin University of Science and Technology, Harbin, 150080, China}

\begin{document}
\maketitle

\begin{abstract}
In this paper we study the weak decays of $J/\psi$ and $\Upsilon(1S)$. Using the Bethe-Salpeter method, we calculate the hadronic transition amplitude and give the form factors. We find that two new form factors $h_1$ and $h_2$, which do not appear in existing literature, have contributions in $1^-\to 1^-$ decays. They affect the branching ratios of semi-leptonic and non-leptonic decays by the rate of $3\%\sim6\%$ and $2\%\sim14\%$, respectively, so their contributions can not be ignored and should be considered. Our results show that, for the semi-leptonic decay modes, the largest branching ratios are of the order of $10^{-10}$ both for $J/\psi$ and $\Upsilon(1S)$ decays, and the largest branching ratios of non-leptonic decays are of the order of $10^{-9}$ for $J/\psi$ and $10^{-10}$ for $\Upsilon(1S)$.

\end{abstract}

\section{Introduction}

Quark potential models have predicted abundant charmonium and bottomonium spectra~\cite{GI}, and lots of them have been discovered experimentally~\cite{PDG}. These states provide ideal laboratories to study non-perturbative properties of QCD. Unlike the open-flavor mesons (e.g.~$B_c$), the $1^{--}$ ground states, namely $J/\psi$ and $\Upsilon(1S)$, which have been produced in large quantities, mainly decay through the strong and electromagnetic processes. These channels have been investigated extensively, while the weak decay channels just attract limited attention~\cite{SL,SR01,SR02,Dhir,Li2,Shen,Sha,Sun01,Sun02,Sun03,Ke,IT}. This is mainly because in the Standard Model (SM), the branching ratios of semi-leptonic and non-leptonic decay channels of both particles are small, and the upper bound are found to be of the order of $10^{-10}\sim 10^{-8}$~\cite{SL}. In PDG~\cite{PDG}, the smallest upper limit of all the experimental results is about $10^{-7}$, which is still much larger than the theoretical prediction of the weak decay channels. Nevertheless, some phenomenological models which are beyond SM, such as two-Higgs-doublet model (2HDM)~\cite{Datta} and top-color model~\cite{Hill}, have predicted that the branching ratios of these channels are enhanced to be around $10^{-5}$.

Although nowadays there is no experimental result to reach the theoretical expectations of the SM, the charm-factory and B-factories are accumulating more and more data, which will bring hope to set more stringent upper limit. For example, the BESIII Collaboration is expected to collect $10^{10}$ $J/\psi$ events per year with the designed luminosity~\cite{LL} (this luminosity has been achieved very recently) and the Belle Collaboration have accumulated over $10^{8}$ $\Upsilon(1S)$ samples~\cite{Bev}. At the LHCb, more than $10^{11}$ $b\bar b$ and $10^{12}$ $c\bar c$ per ${\rm fb}^{-1}$ data are, in principle, available~\cite{LHCb}. The under-constructed new B-factory, namely, SuperKEKB/Belle II, will reach higher luminosity which will exceed Belle by about 40 times, and it will collect more data with great precision~\cite{BelleII}. The future Super-tau-charm Factory is expected to produce $10^{12}$ $J/\psi$ events for half a year at an unprecedentedly high luminosity of $10^{35}~{\rm cm^{-2}s^{-1}}$~\cite{CFT01,CFT02}. Recently, the BESIII Collaboration have published two papers about $J/\psi$ weak decays by using a sample of $2.25\times 10^8$ $J/\psi$ events. Their results are: $Br(J/\psi\rightarrow D_s^-e^+\nu + c.c.)<1.3\times 10^{-6}$, $Br(J/\psi\rightarrow D_s^{\ast-}e^+\nu + c.c.)<1.8\times 10^{-6}$ for the semi-leptonic decays~\cite{BES1} and $Br(J/\psi\rightarrow D_s^-\rho + c.c.)<1.3\times 10^{-5}$, $Br(J/\psi\rightarrow D_s^{\ast-}\rho + c.c.)<2.5\times 10^{-6}$ for non-leptonic decays~\cite{BES2}. Although the upper bound is lowered by more than one order of magnitude than that in Ref~\cite{BES06} (the data sample there is 4 times less), it is still three orders of amplitude larger than the theoretical results~\cite{SR01, SR02}. So more data are needed to achieve the precisions of standard model predictions.

The key point of theoretical calculations for the semi-leptonic decay processes is the hadronic transition matrix element between the initial and final mesons,  which can be characterized by several form factors. Various models have been applied to study the transition form factors, such as the QCD Sum Rules (QCDSR)~\cite{SR01, SR02}, the Bauer-Stech-Wirbel (BSW) model~\cite{Dhir}, covariant light-front quark (CLFQ) model~\cite{Shen}, and very recently the confined covariant quark model (CCQM)~\cite{IT}.  The QCDSR model, as mentioned in Ref.~\cite{IT}, cannot do the calculation in the full range of momentum transfer $Q^2=(P-P_f)^2$, for there are singularities in large $Q^2$. For the other three models, the simple Gaussian-type wave function of the meson is used, while in this paper, we will adopt a more reasonable one, which is a relativistic method and viable in the full range of momentum transfer. The knowledge of form factors is also important for the study of non-leptonic decays, which is only related to one point of the phase space. So these channels depend more sensitively on the form factors.

The non-leptonic decay modes can be divided into several types, however in the existing literature, the authors concentrate only on part of them. In Ref.~\cite{SR02}, the color-favored and Cabibbo-favored (${\rm CoF\otimes CaF}$, see Fig.~1(a)), color-suppressed and Cabibbo-favored (${\rm CoS\otimes CaF}$, see Fig.~1(b)), and color-favored and Cabibbo-suppressed (${\rm CoF\otimes CaS}$, see Fig.~1(a)) channels are studied. Other channels, such as color-suppressed and Cabibbo-suppressed (${\rm CoS\otimes CaS}$) ones, may have decay branching ratios as large as those of the former ones~\cite{Dhir}. However, in Ref.~\cite{Dhir}, the channels with $D^\ast_{(s)}$ as the final meson are not considered. So it is necessary to study these channels overall with a different model.
\begin{figure}[ht]
\centering
\subfigure[]{\includegraphics[scale=0.8]{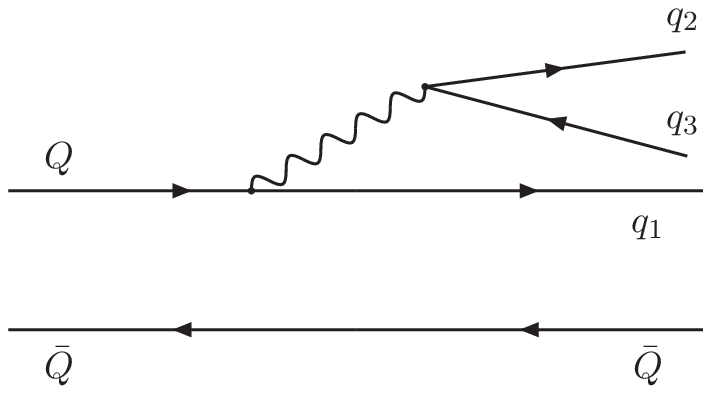}}
\hspace{20 mm}
\subfigure[]{\includegraphics[scale=0.8]{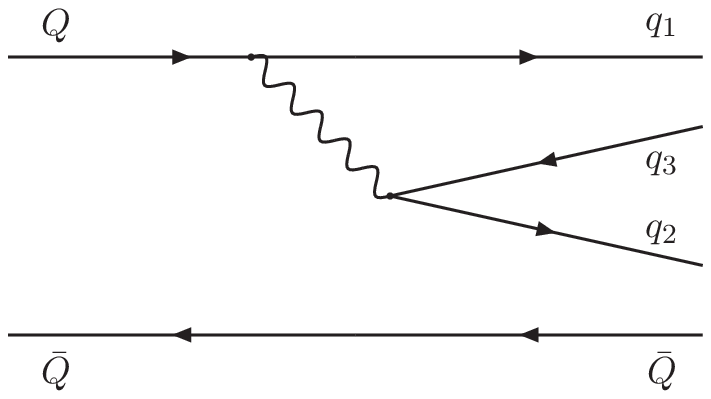}}
\caption[]{Feynman diagrams of non-leptonic decay for (a) the color-allowed case and (b) the color-suppressed case.}
\end{figure}

As the masses of $J/\psi$ and $\Upsilon(1S)$ are much larger than $\Lambda_{QCD}$, the non-relativistic QCD methods are widely applied to study the spectra and decays of these heavy quarkona. However, as Ref.~\cite{SR02} mentioned, they may lose their validity of studying processes involving heavy-light mesons.  Besides that, in the non-leptonic decays of heavy quarkonia, the recoil momenta of final particles may be large enough, so the relativistic effect should be considered~\cite{Shen}. As is known to all, Bethe-Salpeter (BS) equation~\cite{BS1} presents a relativistic description of two-body bound states. By solving its instantaneous form~\cite{BS2}, namely, the full Salpeter equation, one could get the spectrum and wave functions of corresponding states. The transition amplitude is written within Mandelstam formalism~\cite{Man}. This method has been used in our previous papers to study weak decays of heavy-light mesons, such as $B$~\cite{Fu01}, $B^\ast_{sJ}$~\cite{wang2} and $B_c$ states~\cite{Chang}, where we got the results which are in agreement with the experiment data. In this paper, we will use the same method to investigate both the semi-leptonic and non-leptonic decays of $J/\psi$ and $\Upsilon(1S)$.

The paper is organized as follows. In Section 2 we present the theoretical formalism to calculate the semi-leptonic decays of $1^{--}$ heavy quarkonia. Two cases, i.e., the final meson being $0^-$ state or $1^-$ state, are investigated. The non-leptonic decay modes are considered in Section 3. Section 4 is reserved to the results and discussions. Finally, we give our conclusion.

\section{Semi-Leptonic Decay}

The transition amplitude of the semi-leptonic decay processes (see Fig.~2) can be written as the product of the leptonic current and the hadronic matrix element
\begin{equation}
  \mathcal{M}=\frac{G_{F}}{\sqrt{2}}V_{Qq}\bar{u}_{l}\gamma_{\mu}(1-\gamma_{5})v_{\bar{\nu}_{l}}\langle P_f, \epsilon_f|J^{\mu}|P, \epsilon\rangle,
\end{equation}
where $G_F$ is the Fermi coupling constant; $P$, $\epsilon$ and $P_f$, $\epsilon_f$ are the momentum and the polarization vector of initial (with mass M) and final (with mass $M_f$) meson, respectively; $V_{Qq}$ is the CKM matrix element with $Q$ and $q$ representing the initial and final quark, respectively; the transition current is defined as $J^\mu=\bar q \gamma^\mu(1-\gamma_5)Q\equiv V^\mu-A^\mu$.

For the $1^{-}\rightarrow 0^-$ channel, the hadronic matrix element generally has the following form (there is no $\epsilon_f$ in this case)
\begin{equation}
\begin{aligned}
&\langle P_f|V^{\mu}|P, \epsilon\rangle=\frac{2s_1}{M+M_f}i\epsilon^{\mu\nu\rho\sigma}\epsilon_\nu P_\rho P_{f\sigma},\\
&\langle P_f|A^{\mu}|P, \epsilon\rangle= s_2(M+M_f)\epsilon^\mu - (s_3P^\mu - s_4P_f^\mu)\frac{\epsilon\cdot P_f}{M},
\end{aligned}
\end{equation}
where $s_i$s are form factors, and this form is model independent, but the numerical values of form factors are usually model dependent.
As for the $1^{-}\rightarrow 1^-$ channel, the hadronic matrix element is parameterized as
\begin{equation}
\begin{aligned}
&\langle P_f, \epsilon_f|V^{\mu}|P, \epsilon\rangle=(t_1P^\mu + t_2P_f^\mu)\frac{\epsilon\cdot P_f\epsilon_f\cdot P}{M^2}-t_3\epsilon^\mu\epsilon_f\cdot P - t_4\epsilon_f^\mu\epsilon\cdot P_f  \\
&~~~~~~~~~~~~~~~~~~~~~~~+ (t_5P^\mu + t_6P_f^\mu)\epsilon\cdot\epsilon_f,\\
&\langle P_f, \epsilon_f|A^{\mu}|P, \epsilon\rangle=i\epsilon^{\alpha\beta\gamma\delta}\epsilon_\alpha\epsilon_{f\beta} \frac{P_{\gamma} P_{f\delta}}{M^2} (h_1 P^\mu - h_2 P_f^\mu) + i\epsilon^{\mu\alpha\gamma\delta}\frac{P_{\gamma} P_{f\delta} }{M^2}(h_3\epsilon_\alpha\epsilon_f\cdot P \\
&~~~~~~~~~~~~~~~~~~~~~~~- h_4\epsilon_{f\alpha}\epsilon\cdot P_f) + i\epsilon^{\mu\alpha\beta\gamma}\epsilon_{\alpha}\epsilon_{f\beta}(h_5P_\gamma - h_6P_{f\gamma}),
\end{aligned}
\end{equation}
$t_i$s and $h_i$s are the form factors. In the existing literature, the form factors $h_1$ and $h_2$ are missing. The reason is that, in model dependent calculations (non-relativistic or semi-relativistic), $h_1$ and $h_2$ have no contribution, while in our method, they are non-zero.

In our model, the form factors are calculated within Mandelstam formalism~\cite{Man}, that is, the hadronic transition amplitude is written as the overlapping integral of the BS wave functions of initial and final mesons~\cite{max1,max2}:
\begin{equation}
\begin{aligned}
\langle P_f, \epsilon_f|J^{\mu}|P, \epsilon\rangle=\int\frac{d{\vec{q}}}{(2\pi)^{3}}\textrm{Tr}\left[\frac{\rlap{$/$}P}{M}\overline{\varphi_{_{P_f}}^{++}}({\vec{q}_f})\gamma_{\mu}(1-\gamma_{5})\varphi_{_{P}}^{++}({\vec{q}})\right],
\end{aligned}
\end{equation}
where $\vec q$ and $\vec q_f$ are the relative momenta of quark and antiquark in the initial and final mesons, respectively. They are related by $\vec q_f=\vec q-\frac{m^\prime_2}{m_1^\prime+m_2^\prime} \vec P_f$ with $m_1^\prime$ and $m_2^\prime$ being masses of quark and antiquark in the final meson. $\varphi_{P(P_f)}^{++}$ is the positive part of initial (final) meson wave function and $\overline{\varphi_{P_f}^{++}}$ is expressed as $\gamma^0(\varphi_{P_f}^{++})^{\dagger}\gamma^0$. In our calculations, we have ignored the very tiny contributions from other parts of wave functions.

The relativistic wave functions are obtained by solving the full Salpeter equation. But the equation itself does not give the form of the corresponding wave function. So we construct it by $\slashed P$, $\slashed q$, $\slashed\epsilon$, and some scalar functions with the condition that it has the same $J^{P(C)}$ number as that of the corresponding meson. The explicit forms of wave functions and the process of solving corresponding Salpeter equations are given in our former papers~\cite{wang3, wang1}. Here we only give the expressions of vector ($1^-$) and  pseudoscalar ($0^-$) wave functions in Appendix A.

\begin{figure}
\centering
\includegraphics[width=0.5\textwidth]{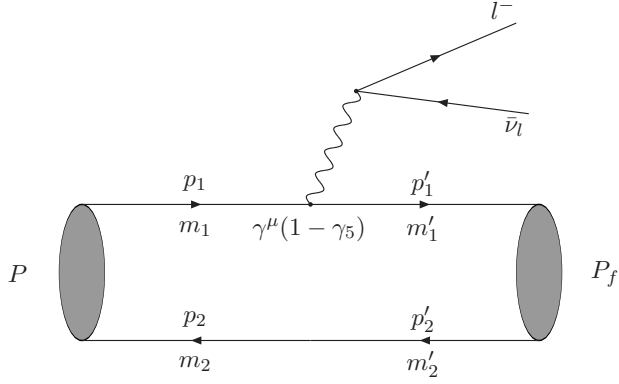}
\caption[]{The Feynman diagram of the semi-leptonic decay processes of $\Upsilon(1S)$. For $J/\psi$, the final leptons is $l^+\nu_l$}
\end{figure}

The decay width has the following form
\begin{equation}
\begin{aligned}
\varGamma=\frac{1}{3}\frac{1}{8M(2\pi)^3}\int dE_{l}dE_{f}\sum_{\lambda}|\mathcal{M}|^{2},
\end{aligned}
\end{equation}
where $E_l$ and $E_f$ represent the energy of the charged lepton and the final meson, respectively. The squared matrix element can be written as the product of hadronic and leptonic tensors
\begin{equation}
  |\mathcal{M}|^{2}=\frac{G_{F}^{2}}{2}|V_{Qq}|^{2}l_{\mu\nu}h^{\mu\nu}.
\end{equation}
The leptonic tensor has the form
\begin{equation}
\begin{aligned}
  l_{\mu\nu}&=\bar{u}_{l}\gamma_{\mu}(1-\gamma_{5})v_{\bar{\nu}_{l}}\bar{v}_{\bar{\nu}_{l}}(1+\gamma_{5})\gamma_{\nu}u_{l}\\
  &=8(p_{l\mu}p_{\bar{\nu}_{l}\nu}+p_{\bar{\nu}_{l}\mu}p_{l\nu}-g_{\mu\nu}p_{l}\cdot p_{\bar{\nu}_{l}}-i\varepsilon_{\mu\nu\rho\sigma}p_{l}^{\rho}p_{\bar{\nu}_{l}}^{\sigma}),
\end{aligned}
\end{equation}
where $p_l$ and $p_{\bar\nu}$ are momenta of the charged lepton and anti-neutrino, respectively (the last term changes sign for the $J/\psi$ case). And the hadronic tensor has the form
\begin{equation}
\begin{aligned}
h^{\mu\nu}=&-\alpha g^{\mu\nu}+\beta_{++}(P+P_{f})^{\mu}(P+P_{f})^{\nu}+\beta_{+-}(P+P_{f})^{\mu}(P-P_{f})^{\nu}\\
&+\beta_{-+}(P-P_{f})^{\mu}(P+P_{f})^{\nu}+\beta_{--}(P-P_{f})^{\mu}(P-P_{f})^{\nu}\\
&+i\gamma\varepsilon^{\mu\nu\rho\sigma}(P+P_{f})_{\rho}(P+P_{f})_{\sigma}.
\end{aligned}
\end{equation}
$\alpha$, $\gamma$, and $\beta_{\pm\pm}$ are functions of form factors, whose explicit expressions are presented in the Appendix B. Finally, the squared amplitude has the form
\begin{equation}
  |\mathcal{M}|^{2}=2G_{F}^{2}|V_{Q\bar q}|^{2}\left[g^{\alpha}\alpha+g^{++}\beta_{++}+g^{+-}(\beta_{+-}+\beta_{-+})+g^{--}\beta_{--}+g^{\gamma}\gamma\right],
\end{equation}
where we have used the following definitions
\begin{equation}
\begin{aligned}
&g^{\alpha}=2M^{2}y-2m_{l}^{2},\\
&g^{++}=4M^{4}\left[2x(1-\frac{M_{f}^{2}}{M^{2}}+y)-4x^{2}-y\right]\\
&~~~~~~~+m_{l}^{2}M^{2}\left(8x-3y+4\frac{M_{f}^{2}}{M^{2}}-\frac{m_{l}^{2}}{M^{2}}\right),\\
&g^{+-}=m_{l}^{2}M^{2}\left(-4x+y+2-2\frac{M_{f}^{2}}{M^{2}}+\frac{m_{l}^{2}}{M^{2}}\right),\\
&g^{--}=m_{l}^{2}M^{2}y-m_{l}^{4},\\
&g^{\gamma}=2M^{4}\left[-4xy+(y+\frac{m_{l}^{2}}{M^{2}})(1-\frac{M_{f}^{2}}{M^{2}}+y)\right].
\end{aligned}
\end{equation}
In the above equations, $x$ and $y$ are defined as
\begin{equation}
\begin{aligned}
&x=\frac{E_{l}}{M},~~~y=\frac{(p_{l}+p_{\bar{\nu}_l})^{2}}{M^{2}}.
\end{aligned}
\end{equation}

\section{Non-Leptonic Decay}

For the non-leptonic decay process, we have integrated out $W^-$, and reduce the vertex to the four-fermion version. The Feynman diagram is given in Figure 3. The Low-energy effective Hamiltonian has the form~\cite{BBL}
\begin{equation}
  \mathcal{H}_{eff}=\frac{G_{F}}{\sqrt{2}}V^{*}_{Qq_1}V_{q_2q_3}\left[C_{1}(\mu)Q_{1}+C_{2}(\mu)Q_{2}\right]+h.c.,
\end{equation}
where $C_{1}(\mu)$ and $C_{2}(\mu)$ are Wilson coefficients and the operators $Q_i$s are defined as
\begin{equation}
\begin{aligned}
  &Q_{1}=({\bar Q}_{\alpha}q_{1\alpha})_{V-A}({\bar q}_{2\beta}q_{3\beta})_{V-A},\\
  \vspace{0.6cm}
  &Q_{2}=({\bar Q}_{\alpha}q_{1\beta})_{V-A}({\bar q}_{2\beta}q_{3\alpha})_{V-A},
\end{aligned}
\end{equation}
where $\alpha$ and $\beta$ represent the color indices. By using the Fierz transformation, we could get~\cite{Choi}
\begin{equation}
\begin{aligned}
C_{1}(\mu)Q_{1}+C_{2}(\mu)Q_{2} = a_1(\mu)Q_{1} + a_2(\mu) Q^{(8)} = a_2(\mu)Q_2 + a_1(\mu) Q^{\prime (8)},
\end{aligned}
\end{equation}
where $Q^{(8)}$ and $Q^{\prime(8)}$ are color-octet current operators which could be neglected in the calculation. And we have defined
\begin{equation}
\begin{aligned}
&a_1(\mu) = C_1(\mu)+ \frac{1}{N_c}C_2(\mu),\\
&a_2(\mu) = C_2(\mu)+ \frac{1}{N_c}C_1(\mu),
\end{aligned}
\end{equation}
which have the values: $a_1(m_c)=1.274$, $a_2(m_c)=-0.529$~\cite{Yu}, and $a_1(m_b)=1.14$, $a_2(m_b)=-0.20$~\cite{Choi}.

The branching ratios of these channels are very small, and we are interested in the order of magnitude. So the nonfactorization contribution can be neglected. By applying the so-called naive-factorization, the transition amplitude $\left\langle P_fP_h|\mathcal{H}_{eff}|J/\psi\right\rangle$ can be written as the product of the hadronic matrix elements (see Eq.~(4)) and the annihilation matrix element, $\left\langle P_h|J_{\mu}|0\right\rangle$, which can be expressed as
\begin{equation}
  \left\langle P|A^{\mu}|0\right\rangle=if_{P}P_P^{\mu}
\end{equation}
and
\begin{equation}
  \left\langle V|V^{\mu}|0\right\rangle=f_{V}m_{_V}\epsilon_{_V}^{\ast\mu}
\end{equation}
for the final state $h$ (see Fig. 3) being a pseudoscalar ($P$) or vector meson ($V$), respectively. In the above equations, $P_P$ and $f_P$ are respectively the momentum and decay constant of the pseudoscalar $h$ meson; $m_V$, $\epsilon_{_V}$, and $f_V$ are respectively the mass, polarization vector, and decay constant of the vector $h$ meson. As Ref.~\cite{Choi} mentioned, for the neutral light meson $\pi^0$, $\rho^0$, or $\omega$, the corresponding decay constant is replaced to $f_{P(V)}/\sqrt{2}$. In Table 1, we present the numerical values of decay constants for different mesons related to this paper.

\begin{figure}
\centering
\includegraphics[width=0.5\textwidth]{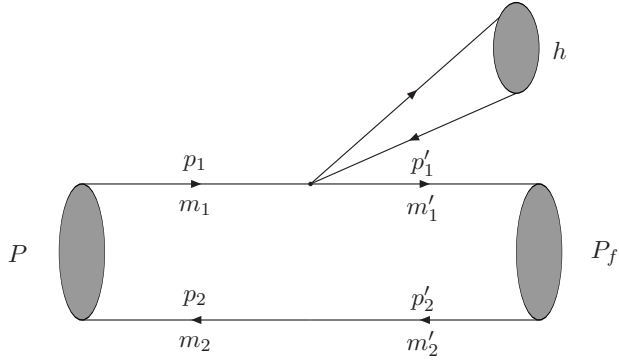}
\caption[]{The Feynman diagram for the non-leptonic decay processes of $J/\psi$ or $\Upsilon(1S)$ with an effective interaction lagrangian.}
\end{figure}

By combining Eq.~(16) (Eq.~(17)) and Eq.~(4), we get the transition amplitude of non-leptonic decays. If $h$ is a pseudoscalar meson, the amplitude has the form
\begin{equation}
  \mathcal{T}=\frac{G_{F}}{\sqrt{2}}V^{*}_{Qq_1}V_{q_2q_3}a_{i}f_{P}P_{P}^{\mu}\langle P_f, \epsilon_f|J^{\mu}|P, \epsilon\rangle.
 \end{equation}
And if $h$ is a vector meson, the amplitude has the form
\begin{equation}
  \mathcal{T}=\frac{G_{F}}{\sqrt{2}}V^{*}_{Qq_1}V_{q_2q_3}a_{i}f_{V}m_{V}\epsilon_{V}^{\ast\mu}\langle P_f, \epsilon_f|J^{\mu}|P, \epsilon\rangle.
\end{equation}
The two-body decay width is written as
\begin{equation}
  \varGamma =\frac{|\vec{P}_{f}|}{8\pi M^{2}}\frac{1}{3}\sum_{\lambda}|\mathcal{T}|^{2}.
\end{equation}

Here we would like to mention the case when the final light meson is $\eta$ or $\eta^\prime$. The two physical states $\eta$ and $\eta^\prime$ are related to the flavor states by~\cite{Yu}
\begin{equation}
\begin{aligned}
&\eta=\frac{1}{\sqrt{2}}(u\bar u + d\bar d )\cos\phi - (s\bar s) \sin\phi,\\
&\eta^\prime = \frac{1}{\sqrt{2}}(u\bar u + d\bar d)\sin\phi + (s\bar s)\cos\phi,
\end{aligned}
\end{equation}
where the mixing angle $\phi=40.4^\circ$ is used.
The decay constants are defined as~\cite{Yu}
\begin{equation}
\begin{aligned}
&\langle0| \bar q\gamma_\mu\gamma_5 q|\eta(P)\rangle = \frac{i}{\sqrt{2}}\cos\phi f_q P_\mu,\\
&\langle0| \bar s\gamma_\mu\gamma_5 s|\eta(P)\rangle = -i\sin\phi f_s P_\mu,\\
&\langle0| \bar q\gamma_\mu\gamma_5 q|\eta^\prime(P)\rangle = \frac{i}{\sqrt{2}}\sin\phi f_q P_\mu,\\
&\langle0| \bar s\gamma_\mu\gamma_5 s|\eta^\prime(P)\rangle = i\cos\phi f_s P_\mu,\\
\end{aligned}
\end{equation}
where we used $f_q=1.07 f_\pi$ and $f_s=1.34 f_\pi$.

\section{Results and Discussions}

To solve the Bethe-Salpeter equation numerically and get the wave functions of initial and final mesons, we use the instantaneous approximation~\cite{BS2}, namely the interaction potential does not depend on the timelike component of the relative momentum. With this approximation the equation is reduced into a three-dimensional form, which is called the full Salpeter equation. In this paper we adopt the screened Cornell potential~\cite{Kim}. It consists of a Coulomb-like and a linear potentials, the former of which comes from the one-gluon exchange and the later of which is supported by the Lattice QCD. The parameters of our model are fixed by fitting the mass spectra, and the interesting reader can find the details of how to solve the full Salperter equations in papers \cite{wang1, Kim},  here we just present the masses of quarks: $m_b=4.96$ GeV, $m_c$ = 1.62 GeV, $m_s$ = 0.5 GeV, $m_u$ = 0.305 GeV, $m_d$ = 0.311 GeV.

The weak decay form factors of $J/\psi$ and $\Upsilon(1S)$, which are functions of $Q^2$, are plotted in Figures~4$\sim$9. For the channels with a pseudoscalar meson as the final state (Fig.~4 and Fig.~5), one can see there are four form factors. For $J/\psi$, the form factors are nearly linear functions, while for $\Upsilon(1S)$, the form factors increase more and more quickly along with $Q^2$.

For the channels with a vector meson as the final state (Figs.~6$\sim$9), there are twelve form factors. Six of them correspond to the vector current transition and six others for the axial current transition. However, in existing literature, there are only four form factors corresponding to the axial current,  but $h_1$ and $h_2$ are not included. Eq.~(3) gives the most general form of the transition amplitude which defines the form factors including $h_1$ and $h_2$. In the non-relativistic or semi-relativistic method, these two form factors have no contributions. While we use the relativistic wave functions, and all the twelve form factors have contributions.  We find that the branching ratios of the semi-leptonic and non-leptonic decays are respectively reduced by $3\sim6\%$ and $2\sim14\%$ if the contributions of $h_1$ and $h_2$ are not included, which means they should be considered.

It is also interesting to calculate the values of form factors at zero momentum transfer $Q^2=0$. In Tables 2$\sim$5, we present our results and other models' (the values are different with the original ones as the definition of form factors is different). For $J/\psi\rightarrow D_{(s)}$ decays, Ref.~\cite{IT}'s results are very close to ours, while other two models got quite different values, especially for $s_4$. This is because, for the last two models, two form factors must be equal at $Q^2=0$ to avoid the divergence~\cite{SR01}. The same thing happens to the $J/\psi\rightarrow D^\ast_{(s)}$ channels, which makes the results in Ref.~\cite{IT} not consistent with ours.

In Figs.~10$\sim$12, we present the energy spectra of the charged lepton in the semi-leptonic decay channels. One notices that the energy spectra of $e^-$ and $\mu^-$ are similar to each other, except that the minimum values of the lepton energy are different. For the $J/\psi$ case (Fig.~10a and Fig.~11a), as the mass of the final meson decreases, the peak value of the spectrum decreases and the position of the peak moves to the right. For the $\Upsilon(1S)$ case, one can see that the spectra are separated into two groups by considering the final mesons: ($B,~B^\ast$) and ($B_c,~B_c^\ast$). In this case, if the charged lepton is $e^-$ (Fig.~10b) or $\mu^-$ (Fig.~11b), along with decreasing of the final meson mass, the peak value increases and the peak position moves to the right.  If the charged lepton is $\tau^-$ (Fig.~12),  for the first group, the peak value of $B_c$ case is smaller than that of $B_c^\ast$ case, while for the second group, the peak value of $B$ case is larger than that of $B^\ast$ case. We also notice that, for the energy spectra of $e^-$, the peak values of the second group are smaller than those of the first group, while for the energy spectra of $\tau^-$, the results are inverse.

\begin{figure}
\centering
\subfigure[]{\includegraphics[scale=0.31]{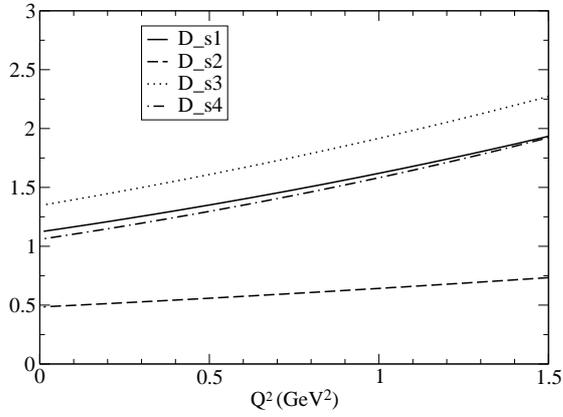}}
\hspace{5 mm}
\subfigure[]{\includegraphics[scale=0.31]{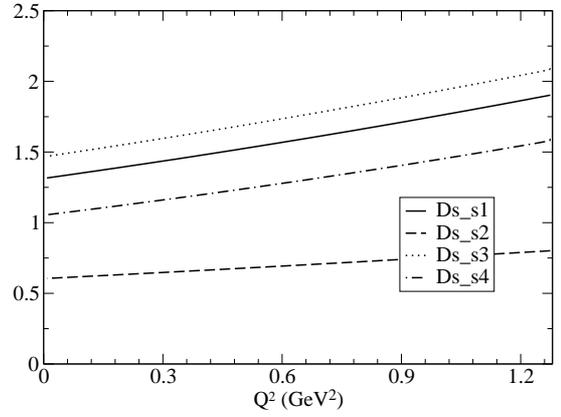}}
\caption[]{Form factors of the transition amplitude for: (a) $J/\psi\rightarrow D^- l^+\nu_l$ (b) $J/\psi\rightarrow D_s^- l^+\nu_l$.}
\end{figure}

\begin{figure}
\centering
\subfigure[]{\includegraphics[scale=0.31]{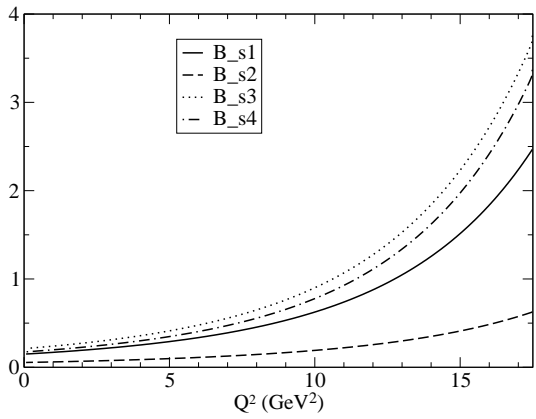}}
\hspace{5 mm}
\subfigure[]{\includegraphics[scale=0.31]{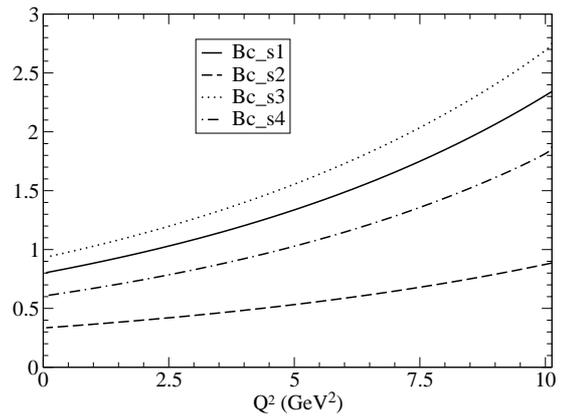}}
\caption[]{Form factors of the transition amplitude for: (a)$\Upsilon\rightarrow B^+l^-\bar\nu_l$ (b) $\Upsilon\rightarrow B_c^+l^-\bar\nu_l$.}
\end{figure}

Our results for the branching ratios of $J/\psi$ semi-leptonic decays are presented in Table~6 (the errors come from the variation of parameters in our model by $\pm$5\%), where those of several other models are also listed. The order of magnitude of the branching ratios is $10^{-11}\sim10^{-10}$. For the channels with $D_{(s)}$ as the final meson, we can see that QCD sum rules~\cite{SR01} give the results which are about half of ours, while the covariant light-front quark model~\cite{Shen}'s results are about two times as large as ours. The BSW model~\cite{Dhir} gets even larger values. The more recently calculation based on the covariant constituent quark model~\cite{IT} give the branching ratios which are very close to ours. For the channels with $D^\ast_{(s)}$ state as the final meson, our results are close to those in Ref.~\cite{SR01} and Ref.~\cite{IT}. To compare the branching ratios of different channels, we calculate $R_i$s which are defined as follows
\begin{equation}
\begin{aligned}
&R^{J/\psi}\equiv \frac{Br[J/\psi\rightarrow D_s^{\ast-}l^+\nu_l]}{Br[J/\psi\rightarrow D_s^-l^+\nu_l]}=1.93~~(l=e,~\mu),\\
&R_1^{J/\psi}\equiv \frac{Br[J/\psi\rightarrow D_s^{-}l^+\nu_l]}{Br[J/\psi\rightarrow D^-l^+\nu_l]}= 18.1~~(l=e,~\mu),\\
&R_2^{J/\psi}\equiv\frac{Br[J/\psi\rightarrow D_s^{\ast-}l^+\nu_l]}{Br[J/\psi\rightarrow D^{\ast-}l^+\nu_l]}=16.1~~(l=e,~\mu).
\end{aligned}
\end{equation}
These ratios are consistent with those in Ref.~\cite{IT}, which have the values: $R=1.5$, $R_1=19.3$, and $R_2=16.6$.

\begin{figure}
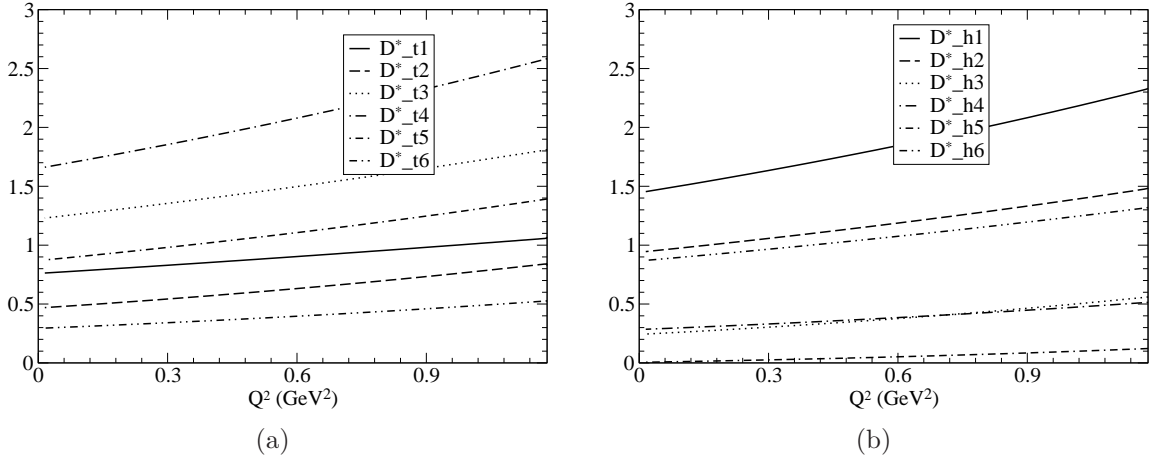

\centering
\subfigure[]{\includegraphics[scale=0.31]{Dstar_t.eps}}
\hspace{5 mm}
\subfigure[]{\includegraphics[scale=0.31]{Dstar_h.eps}}
\caption[]{Form factors of the transition amplitude for the $J/\psi\rightarrow D^{\ast-} l^+\nu_l$ channel: (a) corresponds to the vector current transition, and (b) corresponds to the axial current transition.}
\end{figure}

The semi-leptonic decay results of $\Upsilon(1S)$ are given in Table~7. For the $B_cl\nu_l$ channels, their branching ratios with the BSW model are close to ours. Specifically, when $l=e$ or $\mu$, our results are a little smaller, while for $l=\tau$, our result is larger. We can also give an estimation of $R_i$s for the $\Upsilon(1S)$ semi-leptonic decay channels
\begin{equation}
\begin{aligned}
&R^{\Upsilon(1S)}\equiv\frac{Br[\Upsilon(1S)\rightarrow B_c^{\ast-}l^+\nu_l]}{Br[\Upsilon(1S)\rightarrow B_c^-l^+\nu_l]}=2.66~(l=e,~\mu),~~2.68~(l=\tau),\\
&R_1^{\Upsilon(1S)}\equiv\frac{Br[\Upsilon(1S)\rightarrow B_c^{-}l^+\nu_l]}{Br[\Upsilon(1S)\rightarrow B^-l^+\nu_l]}=195~(l=e,~\mu),~~82.7~(l=\tau),\\
&R_2^{\Upsilon(1S)}\equiv\frac{Br[\Upsilon(1S)\rightarrow B_c^{\ast-} l^+\nu_l]}{Br[\Upsilon(1S)\rightarrow B^{\ast-}l^+\nu_l]}=181~(l=e,~\mu),~~79.4~(l=\tau).
\end{aligned}
\end{equation}
The ratio $R^{\Upsilon(1S)}$ is close to that of $J/\psi$, while $R_1^{\Upsilon(1S)}$ and $R_2^{\Upsilon(1S)}$ are one order of magnitude larger than that of $J/\psi$. In Ref.~\cite{IT}, the authors pointed out that $R_1^{J/\psi}\approx R_2^{J/\psi}\approx |V_{cs}|^2/|V_{cd}|^2$, which implies the $SU(3)_F$ symmetry. For the $\Upsilon(1S)$ case, since $m_c$ is much larger than $m_u$, such symmetry is badly violated, which leads $R_i^{\Upsilon(1S)}(l=e,~\mu)$ deviates a lot from $|V_{cb}|^2/|V_{ub}|^2$. As for $l=\tau$, it has large mass, which makes $R_i$s smaller than those of $e$ and $\mu$.

\begin{figure}
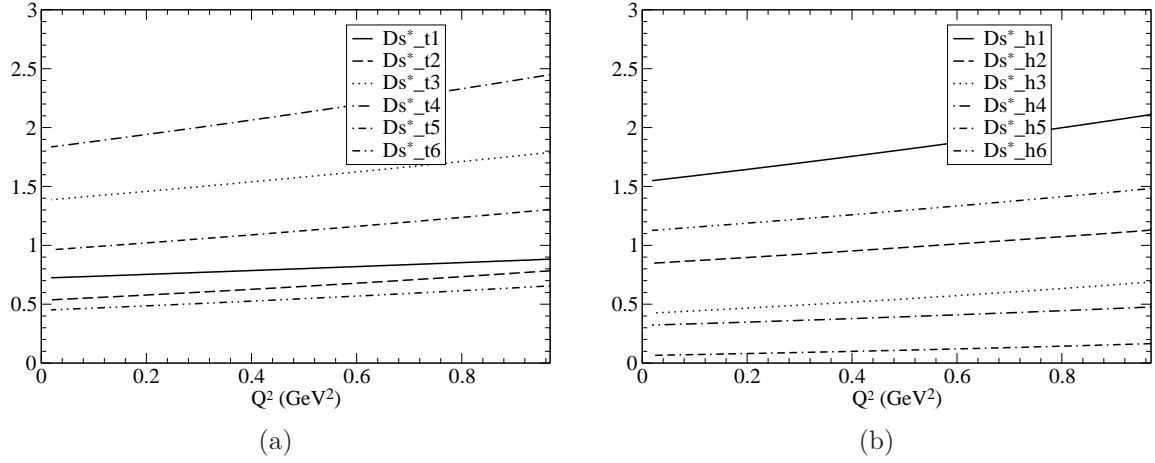

\centering
\subfigure[]{\includegraphics[scale=0.31]{Dsstar_t.eps}}
\hspace{5 mm}
\subfigure[]{\includegraphics[scale=0.31]{Dsstar_h.eps}}
\caption[]{Form factors of the transition amplitude for the $J/\psi\rightarrow D_s^{\ast-} l^+\nu_l$ channel: (a) corresponds to the vector current transition, and (b) corresponds to the axial current transition.}
\end{figure}

\begin{figure}
\centering
\subfigure[]{\includegraphics[scale=0.31]{Bstar_t.eps}}
\hspace{5 mm}
\subfigure[]{\includegraphics[scale=0.31]{Bstar_h.eps}}
\caption[]{Form factors of the transition amplitude for the $\Upsilon\rightarrow B^{\ast+}l^-\bar\nu_l$ channel: (a) corresponds to the vector current transition, and (b) corresponds to the axial current transition.}
\end{figure}

\begin{figure}
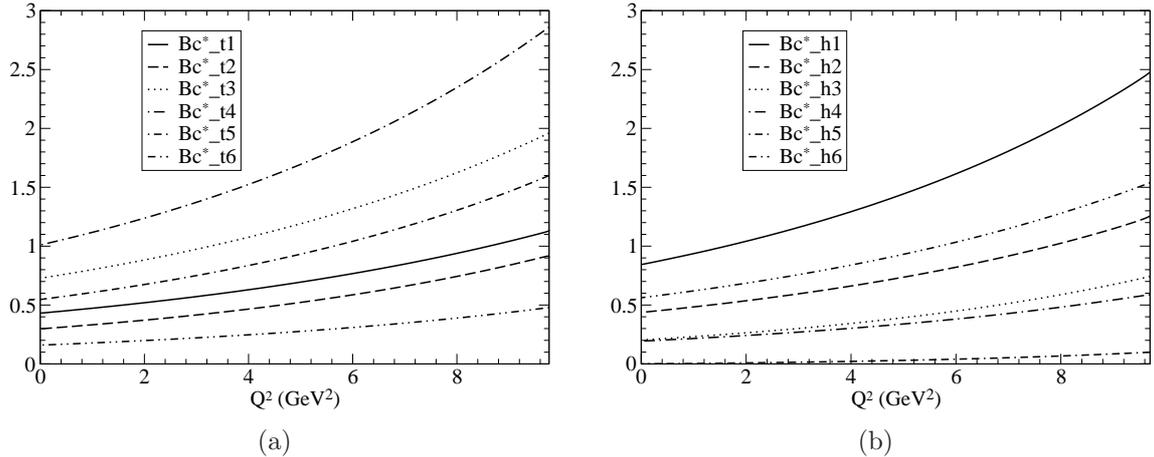

\centering
\subfigure[]{\includegraphics[scale=0.31]{Bcstar_t.eps}}
\hspace{5 mm}
\subfigure[]{\includegraphics[scale=0.31]{Bcstar_h.eps}}
\caption[]{Form factors of the transition amplitude for the $\Upsilon\rightarrow B_c^{\ast+}l^-\bar\nu_l$ channel: (a) corresponds to the vector current transition, and (b) corresponds to the axial current transition.}
\end{figure}

For purposes of comparison, we divide the non-leptonic decay processes of $J/\psi$ into four types: (A) Cabbibo-favored and color-favored (${\rm CaF\otimes CoF}$), (B) Cabibbo-suppressed and color-favored (${\rm CaS\otimes CoF}$, $c_1$) or Cabibbo-favored and color-suppressed (${\rm CaF\otimes CoS}$, $c_2$), (C) Cabibbo-suppressed and color-suppressed (${\rm CaS\otimes CoS}$, $c_3$) or double-Cabibbo-suppressed and color-favored (${\rm CaS^2\otimes CoF}$, $c_4$), (D) double-Cabibbo-suppressed and color-suppressed (${\rm CaS^2\otimes CoS}$) channels. For type A (see Table 8), one can see $D_s^{\ast+}\rho^-+c.c.$ has the largest branching ratio. Our results are larger than those of QCD sum rules~\cite{SR02} but smaller than those of BSW~\cite{Dhir} and QCD factorization~\cite{Sun01} models. The ratio
\begin{equation}
\frac{Br[J/\psi\rightarrow D_s^-\rho^+ +c.c.]}{Br[J/\psi\rightarrow D_s^-\pi^+ +c.c.]} = 5.5
\end{equation}
is close to those of BSW~\cite{Dhir} and QCD sum rules~\cite{SR02}, which are 4.2 and 6.3, respectively. For type B (Table 9), it includes two kinds of decay channels, and each one suffers one kind of suppression, which makes them may have the same order of magnitude. Again, our results are between those of Ref.~\cite{SR02} and Refs.~\cite{Dhir,Sun01}. The channels of type C (Table~10) and type D (Table~11) suffer two and three kinds of suppressions, respectively. Most of them have the branching ratios of the order of $10^{-12}\sim 10^{-11}$.

 One notices that, for channels with $\eta$ or $\eta^\prime$ (Table~10), our results differ greatly from those in Refs.~\cite{Dhir, Sun01}. There are two interference diagrams for these channels. One is $c\rightarrow d\bar du$, and the ohter is $c\rightarrow s\bar s u$. By using Eq.~(22), we can estimate the ratio
 \begin{equation}
 \begin{aligned}
 \frac{Br[J/\psi\rightarrow \overline D^0\eta+ c.c.]}{Br[J/\psi\rightarrow\overline D^0\eta^\prime + c.c.]}\approx\left|\frac{\mathcal K_1P_{\eta\mu}\langle \eta|J^{\mu}|J/\psi\rangle}{\mathcal K_2 P_{\eta^\prime\mu} \langle \eta^\prime|J^{\mu}|J/\psi\rangle}\right|^2\sim\left|\frac{\mathcal K_1}{\mathcal K_2}\right|^2= 0.04,
 \end{aligned}
 \end{equation}
where we have used the definitions $\mathcal K_1 = V_{cd}V_{ud}\cos\phi f_q/\sqrt{2}-V_{cs}V_{us}\sin\phi f_s$ ~and~ $\mathcal K_2 = V_{cd}V_{ud}\sin\phi f_q/\sqrt{2}+V_{cs}V_{us}\cos\phi f_s$. The deviation of this estimation from the actual value, which is 0.07, comes from the mass difference of $\eta$ and $\eta^\prime$ in the hadronic matrix elements.

The non-leptonic decays of $\Upsilon(1S)$ have been classified into two groups: one is the $b\rightarrow u q_2q_3$ decay, and the other is the $b\rightarrow cq_2q_3$ decay. Each group is divided into three types by considering whether the channels are suppressed by flavor and color or not. For the first group, they have small CKM matrix element $V_{ub}$, which makes most of them have decay widths far smaller than those of the second group. The results of the first group are listed in Tables 12$\sim$14. One can see that the smallest ${\rm CaS\otimes CoS}$ channel in Table 14 is of the order of $10^{-18}$, which is far below the current experimental precision. Some ${\rm CaF\otimes CoF}$ decay channels (Table~12) have the branching ratios of the order of $10^{-13}$ which is the same as those of some channels of the second group (see Table 16 and Table 17). The results of the second group are given in Tables 15$\sim$17. We find that the most favorable channel has the branching ratio of the order of $10^{-10}$. Our results of color-favored channels are several times smaller than those in Refs.~\cite{Dhir, Sha} and Refs.~\cite{Sun02,Sun03}. The last two references use the perturbative QCD approach. For the color-suppressed channels, the results in Ref.~\cite{Sha} are much larger than ours.

As a conclusion, we have studied the semi-leptonic and non-leptonic weak decay processes of $J/\psi$ and $\Upsilon(1S)$. We find that two new hadronic transition form factors $h_1$ and $h_2$ have non-zero contributions in the $1^{-}\to1^{-}$ decays. For the semi-leptonic decays, the most prospecting decay channels in experiment are $J/\psi\rightarrow D_s^\ast e\nu_e$ and $\Upsilon(1S)\rightarrow B_c^{\ast}e\nu_e$, whose branching ratios are of the order of $10^{-10}$. The most favorable non-leptonic decay channels are $J/\psi\rightarrow D_s^{\ast-}\rho^++c.c$. and $\Upsilon(1S)\rightarrow B_c^{\ast+}D_s^{\ast-}+c.c.$, which have branching ratios of the order of $10^{-9}$ and $10^{-10}$, respectively. With the existing experiments accumulating more and more data and high-luminosity factories being built in the future, these decay modes may be measured and the theoretical results will be tested.

\begin{figure}
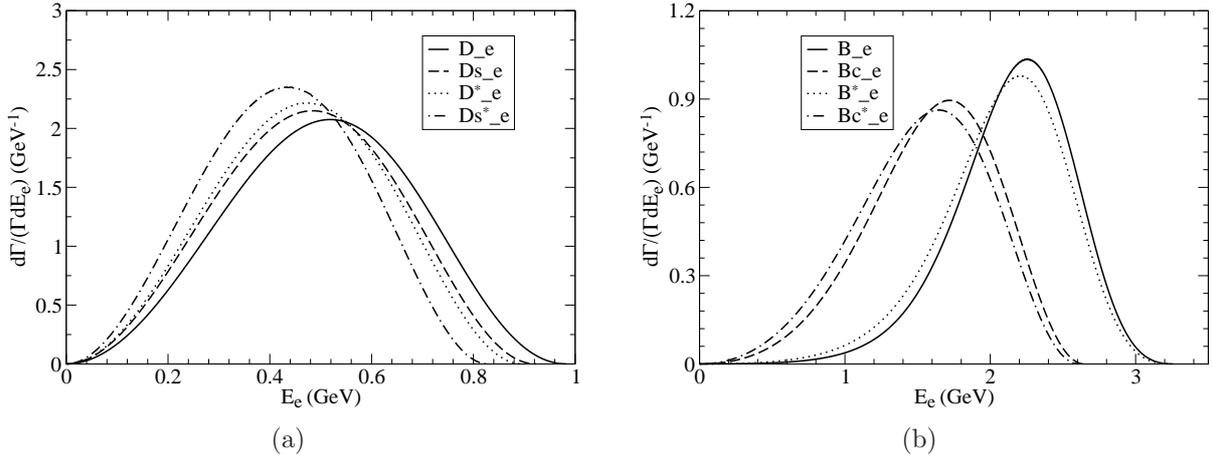

\centering
\subfigure[]{\includegraphics[scale=0.31]{DDs_e.eps}}
\hspace{5 mm}
\subfigure[]{\includegraphics[scale=0.31]{BBc_e.eps}}
\caption[]{Energy spectra of electron for semi-leptonic decays of (a) $J/\psi$ and (b) $\Upsilon(1S)$.}
\end{figure}

\begin{figure}
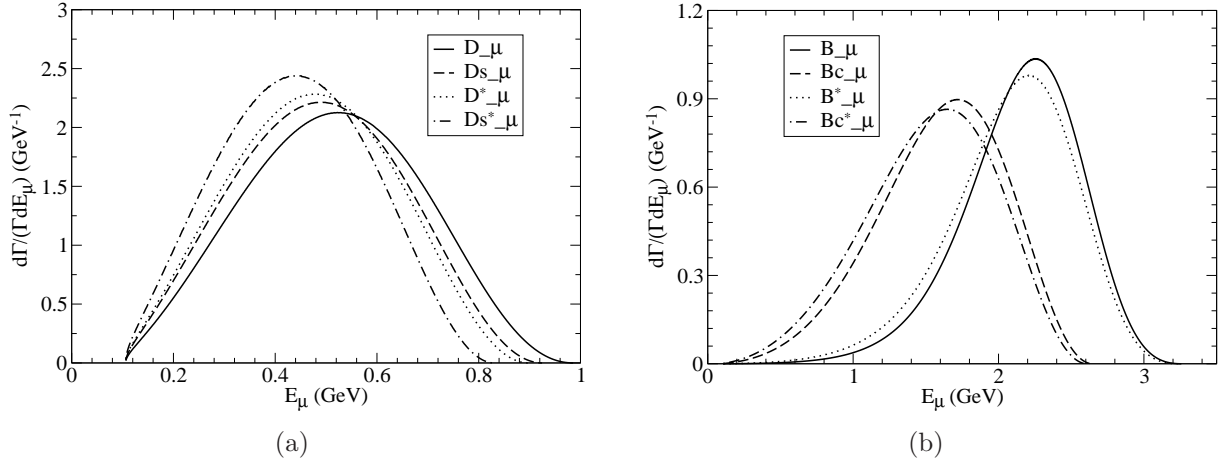

\centering
\subfigure[]{\includegraphics[scale=0.31]{DDs_mu.eps}}
\hspace{5 mm}
\subfigure[]{\includegraphics[scale=0.31]{BBc_mu.eps}}
\caption[]{Energy spectra of muon for semi-leptonic decays of (a) $J/\psi$ and (b) $\Upsilon(1S)$.}
\end{figure}

\begin{figure}
\centering
\includegraphics[scale=0.32]{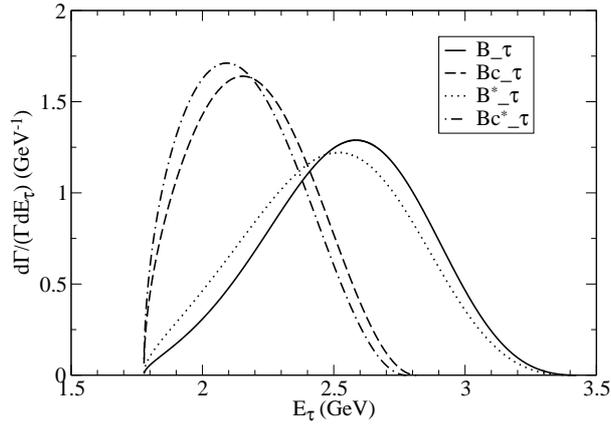}
\caption[]{Energy spectra of tauon for semi-leptonic decays of $\Upsilon(1S)$.}
\end{figure}

\begin{table}[ht]
\caption{Decay constants (MeV) of mesons. Those for $\pi$, $K$, $D$, and $D_s$ are from Particle Data Group~\cite{PDG}; $K^\ast$, $\rho$, $\omega$, and $\phi$ are from Ref.~\cite{Ba}; $D^\ast$ and $D_s^\ast$ are from Ref.~\cite{wang1}; $\eta_c$ and $J/\psi$ are from Ref.~\cite{Bec}.}
\label{results}
\setlength{\tabcolsep}{0.1cm}
\centering
\begin{tabular*}{\textwidth}{@{}@{\extracolsep{\fill}}cccccccccccc}
\hline\hline
{\phantom{\Large{l}}}\raisebox{+.2cm}{\phantom{\Large{j}}}
$f_\pi$&$f_K$&$f_{K^\ast}$&$f_\rho$&$f_\omega$&$f_\phi$&$f_D$&$f_{D_s}$&$f_{D^\ast}$&$f_{D_s^\ast}$&$f_{\eta_c}$&$f_{J/\psi}$\\
{\phantom{\Large{l}}}\raisebox{+.3cm}{\phantom{\Large{j}}}
$130.4$&$156.2$&$217$&$205$&$195$&$231$&$204.6$&$257.5$&$340$&$375$&$387$&$418$\\
\hline\hline
\end{tabular*}
\end{table}

\begin{table}[ht]
 \caption{Form factors at $Q^2=0$ for the $J/\psi\rightarrow D$ processes.}
 \setlength{\tabcolsep}{0.1cm}
 \centering
\begin{tabular*}{\textwidth}{@{}@{\extracolsep{\fill}}ccccc}
\hline \hline
Models &$s_1$&$s_2$&$s_3$ &$s_4$\\ \hline
{\phantom{\Large{l}}}\raisebox{+.2cm}{\phantom{\Large{j}}}
Ours &1.13 &0.48&1.35&1.06\\
{\phantom{\Large{l}}}\raisebox{+.2cm}{\phantom{\Large{j}}}
CCQM\cite{IT} &1.26 &0.44&1.95&1.43\\
{\phantom{\Large{l}}}\raisebox{+.2cm}{\phantom{\Large{j}}}
QCDSR\cite{SR01} & 0.81&0.27&0.21 &-0.21\\
{\phantom{\Large{l}}}\raisebox{+.2cm}{\phantom{\Large{j}}}
LFQM\cite{Shen} &1.6& 0.68&0.11&-0.11\\
\hline\hline
\end{tabular*}
\end{table}

\begin{table}[ht]
 \caption{Form factors at $Q^2=0$ for the $J/\psi\rightarrow D_s$ processes.}
 \setlength{\tabcolsep}{0.1cm}
 \centering
\begin{tabular*}{\textwidth}{@{}@{\extracolsep{\fill}}ccccc}
\hline \hline
Models &$s_1$&$s_2$&$s_3$ &$s_4$\\ \hline
{\phantom{\Large{l}}}\raisebox{+.2cm}{\phantom{\Large{j}}}
Ours & 1.31&0.60&1.47&1.05\\
{\phantom{\Large{l}}}\raisebox{+.2cm}{\phantom{\Large{j}}}
CCQM\cite{IT} & 1.43&0.56&2.07&1.46\\
{\phantom{\Large{l}}}\raisebox{+.2cm}{\phantom{\Large{j}}}
QCDSR\cite{SR01} & 1.07&0.38&0.21 &-0.21\\
{\phantom{\Large{l}}}\raisebox{+.2cm}{\phantom{\Large{j}}}
LFQM\cite{Shen} &1.8&0.68 &0.08&-0.08\\
\hline\hline
\end{tabular*}
\end{table}

\begin{table}[ht]
 \caption{Form factors at $Q^2=0$ for the $J/\psi\rightarrow D^\ast$ processes.}
 \setlength{\tabcolsep}{0.01cm}
 \centering
\begin{tabular*}{\textwidth}{@{}@{\extracolsep{\fill}}ccccccccccccc}
\hline \hline
Models &$t_1$&$t_2$&$t_3$ &$t_4$&$t_5$&$t_6$&$h_1$&$h_2$&$h_3$&$h_4$&$h_5$&$h_6$\\ \hline
{\phantom{\Large{l}}}\raisebox{+.2cm}{\phantom{\Large{j}}}
Ours &0.76 &0.47&1.23&1.66&0.29&0.87&1.45&0.94&0.24&0.28&0.87&0.005\\
{\phantom{\Large{l}}}\raisebox{+.2cm}{\phantom{\Large{j}}}
CCQM\cite{IT} &0.19 &0.19&1.05&1.68&0.12&0.90&---&---&0.71&-0.71&-0.42&0.42\\
\hline\hline
\end{tabular*}
\end{table}

\begin{table}[ht]
 \caption{Form factors at $Q^2=0$ for the $J/\psi\rightarrow D_s^\ast$ processes.}
 \setlength{\tabcolsep}{0.01cm}
 \centering
\begin{tabular*}{\textwidth}{@{}@{\extracolsep{\fill}}ccccccccccccc}
\hline \hline
Models &$t_1$&$t_2$&$t_3$ &$t_4$&$t_5$&$t_6$&$h_1$&$h_2$&$h_3$&$h_4$&$h_5$&$h_6$\\ \hline
{\phantom{\Large{l}}}\raisebox{+.2cm}{\phantom{\Large{j}}}
Ours &0.72 &0.53&1.38&1.83&0.45&0.96&1.54&0.84&0.42&0.32&1.12&0.06\\
{\phantom{\Large{l}}}\raisebox{+.2cm}{\phantom{\Large{j}}}
CCQM\cite{IT} &0.21 &0.21&1.24&1.84&0.26&0.94&---&---&0.69&-0.69&-0.51&0.51\\
\hline\hline
\end{tabular*}
\end{table}

\begin{table}[htb]
 \caption{The branching ratios of semi-leptonic decays of $J/\psi$.}
 \label{semicc}
 \setlength{\tabcolsep}{0.1cm}
 \centering
\begin{tabular*}{\textwidth}{@{}@{\extracolsep{\fill}}ccccccc}
\hline\hline
Channel &Unit&Ours &QCDSR~\cite{SR01}&CLFQ~\cite{Shen}&BSW~\cite{Dhir}&CCQM~\cite{IT}\\ \hline
{\phantom{\Large{l}}}\raisebox{+.2cm}{\phantom{\Large{j}}}
$J/\psi\rightarrow D^{-}e^{+}\nu_{e}$&$10^{-11}$&$2.03^{+0.29}_{-0.25}$&$0.73^{+0.43}_{-0.22}$&$5.1\sim 5.7$& $6.0^{+0.8}_{-0.7}$& $1.71$ \\
{\phantom{\Large{l}}}\raisebox{+.2cm}{\phantom{\Large{j}}}
$J/\psi\rightarrow D^{-}\mu^{+}\nu_{\mu}$&$10^{-11}$&$1.98^{+0.28}_{-0.24}$&$0.71^{+0.42}_{-0.22}$&$4.7\sim 5.5$&$5.8^{+0.8}_{-0.6}$& $1.66$ \\
{\phantom{\Large{l}}}\raisebox{+.2cm}{\phantom{\Large{j}}}
$J/\psi\rightarrow D_{s}^{-}e^{+}\nu_{e}$&$10^{-10}$&$3.67^{+0.52}_{-0.44}$&$1.8^{+0.7}_{-0.5}$&$5.3\sim 5.8$&$10.4^{+0.90}_{-0.75}$& $3.3$ \\
{\phantom{\Large{l}}}\raisebox{+.2cm}{\phantom{\Large{j}}}
$J/\psi\rightarrow D_{s}^{-}\mu^{+}\nu_{\mu}$ & $10^{-10}$&$3.54^{+0.50}_{-0.43}$ &$1.7^{+0.7}_{-0.5}$&$5.5\sim 5.7$&$9.93^{+0.95}_{-0.65}$&$3.2$ \\
{\phantom{\Large{l}}}\raisebox{+.2cm}{\phantom{\Large{j}}}
$J/\psi\rightarrow D^{*-}e^{+}\nu_{e}$ &$10^{-11}$&$4.40^{+0.60}_{-0.52}$&$3.7^{+1.6}_{-1.1}$&& &$3.0$ \\
{\phantom{\Large{l}}}\raisebox{+.2cm}{\phantom{\Large{j}}}
$J/\psi\rightarrow D^{*-}\mu^{+}\nu_{\mu}$&$10^{-11}$ &$4.24^{+0.58}_{-0.50}$&$3.6^{+1.6}_{-1.1}$&& &$2.9$\\
{\phantom{\Large{l}}}\raisebox{+.2cm}{\phantom{\Large{j}}}
$J/\psi\rightarrow D_{s}^{*-}e^{+}\nu_{e}$&$10^{-10}$&$7.08^{+1.14}_{-0.94}$&$5.6^{+1.6}_{-1.6}$&&&$5.0$  \\
{\phantom{\Large{l}}}\raisebox{+.2cm}{\phantom{\Large{j}}}
$J/\psi\rightarrow D_{s}^{*-}\mu^{+}\nu_{\mu}$ &$10^{-10}$ &$6.75^{+1.09}_{-0.90}$&$5.4^{+1.6}_{-1.5}$&& &$4.8$\\ \hline\hline
\end{tabular*}
\end{table}

\begin{table}[htb]
 \caption{The branching ratios of semi-leptonic decays of $\Upsilon(1S)$.}
 \label{semibb}
 \setlength{\tabcolsep}{0.5cm}
 \centering
\begin{tabular*}{\textwidth}{@{}@{\extracolsep{\fill}}cccc}
\hline\hline
Channel &Unit&Ours&BSW~\cite{Dhir} \\ \hline
{\phantom{\Large{l}}}\raisebox{+.2cm}{\phantom{\Large{j}}}
$\Upsilon(1S)\rightarrow B^{+}e^{-}\bar\nu_{e}$  &$10^{-13}$&$7.83^{+1.40}_{-1.20}$ &\\
{\phantom{\Large{l}}}\raisebox{+.2cm}{\phantom{\Large{j}}}
$\Upsilon(1S)\rightarrow B^{+}\mu^{-}\bar\nu_{\mu}$ &$10^{-13}$&$7.82^{+1.40}_{-1.20}$& \\
{\phantom{\Large{l}}}\raisebox{+.2cm}{\phantom{\Large{j}}}
$\Upsilon(1S)\rightarrow B^{+}\tau^{-}\bar\nu_{\tau}$ &$10^{-13}$&$5.04^{+0.92}_{-0.79}$& \\
{\phantom{\Large{l}}}\raisebox{+.2cm}{\phantom{\Large{j}}}
$\Upsilon(1S)\rightarrow B_{c}^{+}e^{-}\bar\nu_{e}$  &$10^{-10}$&$1.37^{+0.22}_{-0.19}$& $1.70^{+0.03}_{-0.02}$ \\
{\phantom{\Large{l}}}\raisebox{+.2cm}{\phantom{\Large{j}}}
$\Upsilon(1S)\rightarrow B_{c}^{+}\mu^{-}\bar\nu_{\mu}$  &$10^{-10}$&$1.37^{+0.22}_{-0.19}$&$1.69^{+0.04}_{-0.02}$ \\
{\phantom{\Large{l}}}\raisebox{+.2cm}{\phantom{\Large{j}}}
$\Upsilon(1S)\rightarrow B_{c}^{+}\tau^{-}\bar\nu_{\tau}$ &$10^{-11}$&$4.17^{+0.58}_{-0.52}$& $2.9^{+0.5}_{-0.2}$ \\
{\phantom{\Large{l}}}\raisebox{+.2cm}{\phantom{\Large{j}}}
$\Upsilon(1S)\rightarrow B^{*+}e^{-}\bar\nu_{e}$ &$10^{-12}$&$2.01^{+0.37}_{-0.32}$& \\
{\phantom{\Large{l}}}\raisebox{+.2cm}{\phantom{\Large{j}}}
$\Upsilon(1S)\rightarrow B^{*+}\mu^{-}\bar\nu_{\mu}$ &$10^{-12}$&$2.01^{+0.37}_{-0.32}$& \\
{\phantom{\Large{l}}}\raisebox{+.2cm}{\phantom{\Large{j}}}
$\Upsilon(1S)\rightarrow B^{*+}\tau^{-}\bar\nu_{\tau}$ &$10^{-12}$&$1.41^{+0.25}_{-0.22}$& \\
{\phantom{\Large{l}}}\raisebox{+.2cm}{\phantom{\Large{j}}}
$\Upsilon(1S)\rightarrow B_{c}^{*+}e^{-}\bar\nu_{e}$&$10^{-10}$ &$3.64^{+0.59}_{-0.51}$& \\
{\phantom{\Large{l}}}\raisebox{+.2cm}{\phantom{\Large{j}}}
$\Upsilon(1S)\rightarrow B_{c}^{*+}\mu^{-}\bar\nu_{\mu}$&$10^{-10}$ &$3.63^{+0.59}_{-0.51}$& \\
{\phantom{\Large{l}}}\raisebox{+.2cm}{\phantom{\Large{j}}}
$\Upsilon(1S)\rightarrow B_{c}^{*+}\tau^{-}\bar\nu_{\tau}$&$10^{-10}$ &$1.12^{+0.16}_{-0.14}$& \\ \hline\hline
\end{tabular*}
\end{table}

\clearpage

\begin{table}[ht]
 \caption{The branching ratios of non-leptonic decays of $J/\psi$ for the Cabibbo-favored and color-favored (${\rm CaF\otimes CoF}$) channels.}
 \setlength{\tabcolsep}{0.01cm}
 \centering
\begin{tabular*}{\textwidth}{@{}@{\extracolsep{\fill}}cccccc}
\hline \hline
Channel &Unit&Ours&QCDSR\cite{SR02} &BSW\cite{Dhir}&QCDF\cite{Sun01} \\ \hline
{\phantom{\Large{l}}}\raisebox{+.2cm}{\phantom{\Large{j}}}
$J/\psi\rightarrow D_{s}^{-}\pi^{+}+D_{s}^{+}\pi^{-}$   &$10^{-10}$&$4.75^{+0.67}_{-0.59}$ &$2.0^{+0.4}_{-0.2}$&$14.82^{+0.26}_{-0.46}$ &21.8\\
{\phantom{\Large{l}}}\raisebox{+.2cm}{\phantom{\Large{j}}}
$J/\psi\rightarrow D_{s}^{-}\rho^{+}+D_{s}^{+}\rho^{-}$  &$10^{-9}$&$2.62^{+0.37}_{-0.32}$ &$1.26^{+0.3}_{-0.1}$&$10.20^{+1.52}_{-1.20}$ &7.64\\
{\phantom{\Large{l}}}\raisebox{+.2cm}{\phantom{\Large{j}}}
$J/\psi\rightarrow D_{s}^{*-}\pi^{+}+D_{s}^{*+}\pi^{-}$   &$10^{-9}$ &$2.57^{+0.34}_{-0.31}$ &$1.5^{+0.12}_{-0.04}$& &\\
{\phantom{\Large{l}}}\raisebox{+.2cm}{\phantom{\Large{j}}}
$J/\psi\rightarrow D_{s}^{*-}\rho^{+}+D_{s}^{*+}\rho^{-}$ & $10^{-9}$&$5.86^{+0.78}_{-0.67}$ &$5.26^{+0.72}_{-0.62}$& &\\ \hline\hline
\end{tabular*}
\end{table}

\begin{table}[ht]
 \caption{The branching ratios of non-leptonic decays of $J/\psi$ for the Cabibbo-Suppressed and color-favored (${\rm CaS\otimes CoF}$, labeled by $c_1$) or Cabibbo-favored and color-suppressed (${\rm CaF\otimes CoS}$, labeled by $c_2$) channels.}
 \setlength{\tabcolsep}{0.1cm}
 \centering
\begin{tabular*}{\textwidth}{@{}@{\extracolsep{\fill}}ccccccc}
\hline \hline
Channel &Type&Unit&Ours&QCDSR\cite{SR02} &BSW~\cite{Dhir}& QCDF\cite{Sun01} \\ \hline
{\phantom{\Large{l}}}\raisebox{+.2cm}{\phantom{\Large{j}}}
$J/\psi\rightarrow D^{-}\pi^{+}+D^{+}\pi^{-}$ &$c_1$ & $10^{-11}$&$1.83^{+0.27}_{-0.25}$ &$0.80^{+0.20}_{-0.20}$& $5.8^{+0.4}_{-0.6}$ &12.7\\
{\phantom{\Large{l}}}\raisebox{+.2cm}{\phantom{\Large{j}}}
$J/\psi\rightarrow D^{0}K^{0}+\overline D^{0}\overline K^{0}$ & $c_2$ &$10^{-11}$&$8.03^{+1.13}_{-1.03}$  &$3.6^{+1.0}_{-0.8}$& $27.8^{+2.0}_{-2.8}$ & 28.8\\
{\phantom{\Large{l}}}\raisebox{+.2cm}{\phantom{\Large{j}}}
$J/\psi\rightarrow D^{0}K^{*0}+\overline D^{0}\overline K^{*0}$& $c_2$ &$10^{-10}$&$4.75^{+0.68}_{-0.58}$ &$1.54^{+0.68}_{-0.38}$&$15.2^{+3.2}_{-2.4}$ &8.18\\
{\phantom{\Large{l}}}\raisebox{+.2cm}{\phantom{\Large{j}}}
$J/\psi\rightarrow D^{-}\rho^{+}+D^{+}\rho^{-}$& $c_1$ &$10^{-10}$&$1.13^{+0.16}_{-0.14}$ &$0.42^{+0.18}_{-0.08}$&$4.32^{+1.0}_{-0.6}$ &4.24\\
{\phantom{\Large{l}}}\raisebox{+.2cm}{\phantom{\Large{j}}}
$J/\psi\rightarrow D_{s}^{-}K^{+}+D_{s}^{+}K^{-}$&$c_1$  &$10^{-11}$&$3.12^{+0.42}_{-0.36}$&$1.6^{+0.2}_{-0.2}$& $10.6^{+0.4}_{-0.4}$  &12.4\\
{\phantom{\Large{l}}}\raisebox{+.2cm}{\phantom{\Large{j}}}
$J/\psi\rightarrow D_{s}^{-}K^{*+}+D_{s}^{+}K^{*-}$ & $c_1$  & $10^{-10}$&$1.67^{+0.24}_{-0.20}$ &$0.82^{+0.22}_{-0.10}$& $5.64^{+0.80}_{-0.60}$&4.0\\
{\phantom{\Large{l}}}\raisebox{+.2cm}{\phantom{\Large{j}}}
$J/\psi\rightarrow D^{*-}\pi^{+}+D^{*+}\pi^{-}$ & $c_1$ &$10^{-10}$ &$1.16^{+0.17}_{-0.15}$ &$0.60^{+0.04}_{-0.04}$& &\\
{\phantom{\Large{l}}}\raisebox{+.2cm}{\phantom{\Large{j}}}
$J/\psi\rightarrow D^{*0}K^{0}+\overline D^{*0}\overline K^{0}$&  $c_2$ &$10^{-10}$ &$5.37^{+0.76}_{-0.68}$&$2.6^{+0.2}_{-0.2}$& & \\
{\phantom{\Large{l}}}\raisebox{+.2cm}{\phantom{\Large{j}}}
$J/\psi\rightarrow D^{*0}K^{*0}+\overline D^{*0}\overline K^{*0}$&$c_2$ &$10^{-9}$  &$1.11^{+0.15}_{-0.13}$ &$0.96^{+0.32}_{-0.22}$& & \\
{\phantom{\Large{l}}}\raisebox{+.2cm}{\phantom{\Large{j}}}
$J/\psi\rightarrow D^{*-}\rho^{+}+D^{*+}\rho^{-}$& $c_1$&$10^{-10}$&$3.30^{+0.45}_{-0.39}$ &$2.8^{+0.6}_{-0.4}$& &\\
{\phantom{\Large{l}}}\raisebox{+.2cm}{\phantom{\Large{j}}}
$J/\psi\rightarrow D_{s}^{*-}K^{+}+D_{s}^{*+}K^{-}$&  $c_1$ &$10^{-10}$ &$1.79^{+0.23}_{-0.20}$  &$1.1^{+0.08}_{-0.04}$& &\\
{\phantom{\Large{l}}}\raisebox{+.2cm}{\phantom{\Large{j}}}
$J/\psi\rightarrow D_{s}^{*-}K^{*+}+D_{s}^{*+}K^{*-}$ &$c_1$ &$10^{-10}$  &$2.62^{+0.35}_{-0.30}$    &$2.6^{+0.4}_{-0.4}$& &\\ \hline\hline
\end{tabular*}
\end{table}

\begin{table}[ht]
 \caption{The branching ratios of non-leptonic decays of $J/\psi$ for the Cabibbo-suppressed and color-suppressed (${\rm CaS\otimes CoS}$, labeled by $c_3$) or double-Cabibbo-suppressed and color-favored (${\rm CaS^2\otimes CoF}$, labeled by $c_4$) channels.}
 \setlength{\tabcolsep}{0.1cm}
 \centering
\begin{tabular*}{\textwidth}{@{}@{\extracolsep{\fill}}cccccc}
\hline \hline
Channel &Type&Unit&Ours&BSW~\cite{Dhir}&QCDF\cite{Sun01} \\ \hline
{\phantom{\Large{l}}}\raisebox{+.2cm}{\phantom{\Large{j}}}
$J/\psi\rightarrow \overline D^{0}\pi^{0}+D^{0}\pi^{0}$ & $c_3$&$10^{-12}$ & $1.56^{+0.24}_{-0.21}$&$4.8^{+0.4}_{-0.6}$ &7.0 \\
{\phantom{\Large{l}}}\raisebox{+.2cm}{\phantom{\Large{j}}}
$J/\psi\rightarrow \overline D^{0}\eta+D^{0}\eta$&$c_3$ & $10^{-13}$&$2.63^{+0.37}_{-0.33}$& $140^{+10}_{-14}$&206 \\
{\phantom{\Large{l}}}\raisebox{+.2cm}{\phantom{\Large{j}}}
$J/\psi\rightarrow \overline D^{0}\eta^{\prime}+D^{0}\eta^{\prime}$&$c_3$& $10^{-12}$ &$3.71^{+0.49}_{-0.42}$& $0.80^{+0.04}_{-0.10}$ &1.17\\
{\phantom{\Large{l}}}\raisebox{+.2cm}{\phantom{\Large{j}}}
$J/\psi\rightarrow D^{-}K^{+}+ D^{+} K^{-}$ &$c_4$&$10^{-12}$&$1.31^{+0.19}_{-0.17}$&$4.6^{+0.4}_{-0.4}$  &7.58\\
{\phantom{\Large{l}}}\raisebox{+.2cm}{\phantom{\Large{j}}}
$J/\psi\rightarrow D^{-}K^{*+}+ D^{+} K^{*-}$& $c_4$&$10^{-12}$ &$7.70^{+1.10}_{-0.93}$&$26^{+4}_{-6}$ &22.8\\
{\phantom{\Large{l}}}\raisebox{+.2cm}{\phantom{\Large{j}}}
$J/\psi\rightarrow \overline D^{0}\rho^{0}+D^{0}\rho^{0}$&$c_3$& $10^{-12}$&$9.60^{+1.35}_{-1.20}$&$36^{+6}_{-6}$ &21.6\\
{\phantom{\Large{l}}}\raisebox{+.2cm}{\phantom{\Large{j}}}
$J/\psi\rightarrow \overline D^{0}\omega^{0}+D^{0}\omega^{0}$&$c_3$&$10^{-12}$&$8.80^{+1.25}_{-1.10}$&$32^{+6}_{-4}$ &16.2\\
{\phantom{\Large{l}}}\raisebox{+.2cm}{\phantom{\Large{j}}}
$J/\psi\rightarrow \overline D^{\ast0}\pi^{0}+D^{\ast0}\pi^{0}$& $c_3$ & $10^{-12}$&$9.90^{+1.50}_{-1.30}$& & \\
{\phantom{\Large{l}}}\raisebox{+.2cm}{\phantom{\Large{j}}}
$J/\psi\rightarrow \overline D^{\ast0}\eta+D^{\ast0}\eta$&$c_3$ & $10^{-12}$ &$1.78^{+0.25}_{-0.22}$ &  &\\
{\phantom{\Large{l}}}\raisebox{+.2cm}{\phantom{\Large{j}}}
$J/\psi\rightarrow \overline D^{\ast0}\eta^{\prime}+D^{\ast}\eta^{\prime}$ &$c_3$& $10^{-11}$ &$2.98^{+0.42}_{-0.36}$ & & \\
{\phantom{\Large{l}}}\raisebox{+.2cm}{\phantom{\Large{j}}}
$J/\psi\rightarrow D^{\ast-}K^{+}+ D^{\ast+} K^{-}$ & $c_4$&$10^{-12}$ & $8.73^{+1.24}_{-1.10}$& & \\
{\phantom{\Large{l}}}\raisebox{+.2cm}{\phantom{\Large{j}}}
$J/\psi\rightarrow D^{\ast-}K^{*+}+ D^{\ast+} K^{*+}$ &$c_4$&$10^{-11}$ &$1.80^{+0.25}_{-0.21}$& &\\
{\phantom{\Large{l}}}\raisebox{+.2cm}{\phantom{\Large{j}}}
$J/\psi\rightarrow \overline D^{\ast0}\rho^{0}+D^{\ast0}\rho^{0}$ &$c_3$&$10^{-11}$ &$2.84^{+0.39}_{-0.34}$ & &\\
{\phantom{\Large{l}}}\raisebox{+.2cm}{\phantom{\Large{j}}}
$J/\psi\rightarrow \overline D^{\ast0}\omega^{0}+D^{\ast0}\omega^{0}$&$c_3$ & $10^{-11}$&$2.55^{+0.35}_{-0.31}$ && \\
{\phantom{\Large{l}}}\raisebox{+.2cm}{\phantom{\Large{j}}}
$J/\psi\rightarrow \overline D^{0}\phi+ D^{0}\phi$& $c_3$&$10^{-11}$ &$3.07^{+0.45}_{-0.38}$ & $8.4^{+1.6}_{-1.4}$&3.84 \\
{\phantom{\Large{l}}}\raisebox{+.2cm}{\phantom{\Large{j}}}
$J/\psi\rightarrow \overline D^{\ast0}\phi+ D^{\ast0}\phi$&$c_3$ &$10^{-11}$ &$4.62^{+0.66}_{-0.56}$ & \\
\hline\hline
\end{tabular*}
\end{table}

\begin{table}[ht]
 \caption{The branching ratios of non-leptonic decays of $J/\psi$ for the Double-Cabibbo-suppressed and color-suppressed (${\rm CaS^2\otimes CoS}$) channels.}
 \setlength{\tabcolsep}{0.1cm}
 \centering
\begin{tabular*}{\textwidth}{@{}@{\extracolsep{\fill}}ccccc}
\hline \hline
Channel &Unit&Ours&BSW~\cite{Dhir} &QCDF\cite{Sun01}\\ \hline
{\phantom{\Large{l}}}\raisebox{+.2cm}{\phantom{\Large{j}}}
$J/\psi\rightarrow \overline D^{0}K^{0}+D^{0}\overline K^{0}$ &$10^{-13}$ &$2.24^{+0.32}_{-0.29}$&$8^{+0}_{-2}$ &8.32\\
{\phantom{\Large{l}}}\raisebox{+.2cm}{\phantom{\Large{j}}}
$J/\psi\rightarrow \overline D^{0}K^{\ast0}+D^{0}\overline K^{\ast0}$ &$10^{-12}$ & $1.32^{+0.19}_{-0.16}$&$4.2^{+0.8}_{-0.6}$&2.38\\
{\phantom{\Large{l}}}\raisebox{+.2cm}{\phantom{\Large{j}}}
$J/\psi\rightarrow \overline D^{\ast0}K^{0}+D^{\ast0}\overline K^{0}$ &$10^{-12}$ &$1.49^{+0.21}_{-0.19}$& &\\
{\phantom{\Large{l}}}\raisebox{+.2cm}{\phantom{\Large{j}}}
$J/\psi\rightarrow \overline D^{\ast0}K^{\ast0}+D^{\ast0}\overline K^{\ast0}$ & $10^{-12}$& $3.09^{+0.43}_{-0.37}$&&\\
\hline\hline
\end{tabular*}
\end{table}

\begin{table}[ht]
 \caption{The branching ratios of non-leptonic decays of $\Upsilon(1S)$ for the Cabibbo-favored and color-favored (${\rm CaF\otimes CoF}$) $b\rightarrow uq_2\bar q_3$ channels.}
 \setlength{\tabcolsep}{0.5cm}
 \centering
\begin{tabular*}{\textwidth}{@{}@{\extracolsep{\fill}}ccc}
\hline \hline
Channel &Unit&Ours \\ \hline
{\phantom{\Large{l}}}\raisebox{+.2cm}{\phantom{\Large{j}}}
$\Upsilon(1S)\rightarrow B^{+}\pi^{-}+B^{-}\pi^{+}$ &$10^{-15}$ &$3.91^{+0.89}_{-0.80}$\\
{\phantom{\Large{l}}}\raisebox{+.2cm}{\phantom{\Large{j}}}
$\Upsilon(1S)\rightarrow B^{*+}\pi^{-}+B^{*-}\pi^{+}$  &$10^{-14}$ &$2.58^{+0.53}_{-0.50}$\\
{\phantom{\Large{l}}}\raisebox{+.2cm}{\phantom{\Large{j}}}
$\Upsilon(1S)\rightarrow B^{+}\rho^{-}+B^{-}\rho^{+}$ & $10^{-14}$&$1.24^{+0.27}_{-0.24}$\\
{\phantom{\Large{l}}}\raisebox{+.2cm}{\phantom{\Large{j}}}
$\Upsilon(1S)\rightarrow B^{*+}\rho^{-}+B^{*-}\rho^{+}$ &$10^{-14}$&$7.47^{+1.47}_{-1.40}$\\
{\phantom{\Large{l}}}\raisebox{+.2cm}{\phantom{\Large{j}}}
$\Upsilon(1S)\rightarrow B^{+}D_s^{-}+B^{-}D_s^{+}$&$10^{-14}$ &$3.05^{+0.58}_{-0.54}$\\
{\phantom{\Large{l}}}\raisebox{+.2cm}{\phantom{\Large{j}}}
$\Upsilon(1S)\rightarrow B^{+}D_s^{*-}+B^{-}D_s^{*+}$ &$10^{-13}$ &$1.62^{+0.33}_{-0.28}$\\
{\phantom{\Large{l}}}\raisebox{+.2cm}{\phantom{\Large{j}}}
$\Upsilon(1S)\rightarrow B^{*+}D_s^{-}+B^{*-}D_s^{+}$ &$10^{-13}$ &$2.17^{+0.42}_{-0.37}$\\
{\phantom{\Large{l}}}\raisebox{+.2cm}{\phantom{\Large{j}}}
$\Upsilon(1S)\rightarrow B^{*+}D_s^{*-}+B^{*-}D_s^{*+}$ & $10^{-13}$ &$6.85^{+1.37}_{-1.20}$\\
\hline\hline
\end{tabular*}
\end{table}

\begin{table}[ht]
 \caption{The branching ratios of non-leptonic decays of $\Upsilon(1S)$ for the Cabibbo-suppressed and color-favored (${\rm CaS\otimes CoF}$, labeled by $b_1$) or Cabibbo-favored and color-suppressed (${\rm CaF\otimes CoS}$, labeled by $b_2$) $b\rightarrow uq_2\bar q_3$ channels.}
 \setlength{\tabcolsep}{0.7cm}
 \centering
\begin{tabular*}{\textwidth}{@{}@{\extracolsep{\fill}}cccc}
\hline \hline
Channel &Type&Unit&Ours \\ \hline
{\phantom{\Large{l}}}\raisebox{+.2cm}{\phantom{\Large{j}}}
$\Upsilon(1S)\rightarrow B^{0}\pi^{0}+\overline B^{0}\pi^{0}$ & $b_2$&$10^{-17}$&$6.15^{+1.40}_{-1.25}$ \\
{\phantom{\Large{l}}}\raisebox{+.2cm}{\phantom{\Large{j}}}
$\Upsilon(1S)\rightarrow B^{\ast0}\pi^{0}+\overline B^{\ast0}\pi^{0}$ & $b_2$&$10^{-16}$&$4.04^{+0.84}_{-0.74}$ \\
{\phantom{\Large{l}}}\raisebox{+.2cm}{\phantom{\Large{j}}}
$\Upsilon(1S)\rightarrow B^{0}\eta+\overline B^{0}\eta$&$b_2$ &$10^{-17}$ &$4.28^{+0.96}_{-0.87}$ \\
{\phantom{\Large{l}}}\raisebox{+.2cm}{\phantom{\Large{j}}}
$\Upsilon(1S)\rightarrow B^{\ast0}\eta+\overline B^{\ast0}\eta$& $b_2$&$10^{-16}$ &$2.83^{+0.58}_{-0.53}$ \\
{\phantom{\Large{l}}}\raisebox{+.2cm}{\phantom{\Large{j}}}
$\Upsilon(1S)\rightarrow B^{0}\eta^{\prime}+\overline B^{0}\eta^{\prime}$& $b_2$& $10^{-17}$&$3.45^{+0.75}_{-0.69}$ \\
{\phantom{\Large{l}}}\raisebox{+.2cm}{\phantom{\Large{j}}}
$\Upsilon(1S)\rightarrow B^{\ast0}\eta^{\prime}+\overline B^{\ast}\eta^{\prime}$& $b_2$& $10^{-16}$&$2.31^{+0.46}_{-0.42}$ \\
{\phantom{\Large{l}}}\raisebox{+.2cm}{\phantom{\Large{j}}}
$\Upsilon(1S)\rightarrow B^{0}\rho^{0}+\overline B^{0}\rho^{0}$&$b_2$ &$10^{-16}$ &$1.94^{+0.42}_{-0.38}$ \\
{\phantom{\Large{l}}}\raisebox{+.2cm}{\phantom{\Large{j}}}
$\Upsilon(1S)\rightarrow B^{\ast0}\rho^{0}+\overline B^{\ast0}\rho^{0}$&$b_2$ & $10^{-15}$&$1.17^{+0.24}_{-0.22}$ \\
{\phantom{\Large{l}}}\raisebox{+.2cm}{\phantom{\Large{j}}}
$\Upsilon(1S)\rightarrow B^{0}\omega^{0}+\overline B^{0}\omega^{0}$&$b_2$ &$10^{-16}$ &$1.77^{+0.38}_{-0.35}$ \\
{\phantom{\Large{l}}}\raisebox{+.2cm}{\phantom{\Large{j}}}
$\Upsilon(1S)\rightarrow B^{\ast0}\omega^{0}+\overline B^{\ast0}\omega^{0}$&$b_2$  &$10^{-15}$&$1.06^{+0.22}_{-0.20}$ \\
{\phantom{\Large{l}}}\raisebox{+.2cm}{\phantom{\Large{j}}}
$\Upsilon(1S)\rightarrow B^{+}K^{-}+B^{-}K^{+}$&$b_1$ &$10^{-16}$&$3.12^{+0.70}_{-0.63}$ \\
{\phantom{\Large{l}}}\raisebox{+.2cm}{\phantom{\Large{j}}}
$\Upsilon(1S)\rightarrow B^{+}K^{*-}+B^{-}K^{*+}$&$b_1$ &$10^{-16}$ &$8.01^{+1.70}_{-1.56}$ \\
{\phantom{\Large{l}}}\raisebox{+.2cm}{\phantom{\Large{j}}}
$\Upsilon(1S)\rightarrow B^{*+}K^{-}+B^{*-}K^{+}$&$b_1$ &$10^{-15}$ &$2.07^{+0.42}_{-0.39}$ \\
{\phantom{\Large{l}}}\raisebox{+.2cm}{\phantom{\Large{j}}}
$\Upsilon(1S)\rightarrow B^{*+}K^{*-}+B^{*-}K^{*+}$& $b_1$ &$10^{-15}$ &$4.71^{+0.93}_{-0.87}$ \\
{\phantom{\Large{l}}}\raisebox{+.2cm}{\phantom{\Large{j}}}
$\Upsilon(1S)\rightarrow B^{+}D^{-}+B^{-}D^{+}$ &$b_1$&$10^{-16}$ &$9.41^{+1.79}_{-1.68}$ \\
{\phantom{\Large{l}}}\raisebox{+.2cm}{\phantom{\Large{j}}}
$\Upsilon(1S)\rightarrow B^{+}D^{\ast-}+B^{-}D^{\ast+}$& $b_1$& $10^{-15}$&$6.08^{+1.24}_{-1.06}$ \\
{\phantom{\Large{l}}}\raisebox{+.2cm}{\phantom{\Large{j}}}
$\Upsilon(1S)\rightarrow B^{*+}D^{-}+B^{*-}D^{+}$ & $b_1$&$10^{-15}$ & $6.64^{+1.29}_{-1.14}$\\
{\phantom{\Large{l}}}\raisebox{+.2cm}{\phantom{\Large{j}}}
$\Upsilon(1S)\rightarrow B^{*+}D^{\ast-}+B^{*-}D^{\ast+}$&$b_1$&$10^{-14}$ &$2.64^{+0.53}_{-0.46}$\\
{\phantom{\Large{l}}}\raisebox{+.2cm}{\phantom{\Large{j}}}
$\Upsilon(1S)\rightarrow B_s^{0}\overline D^{0}+\overline B_s^{0}D^{0}$&$b_2$&$10^{-15}$ &$1.04^{+0.23}_{-0.19}$\\
{\phantom{\Large{l}}}\raisebox{+.2cm}{\phantom{\Large{j}}}
$\Upsilon(1S)\rightarrow B_s^{0}\overline D^{\ast0}+\overline B_s^{0}D^{\ast0}$&$b_2$&$10^{-15}$ &$6.74^{+1.54}_{-1.28}$\\
{\phantom{\Large{l}}}\raisebox{+.2cm}{\phantom{\Large{j}}}
$\Upsilon(1S)\rightarrow B_s^{\ast0}\overline D^{0}+\overline B_s^{\ast0}D^{0}$&$b_2$&$10^{-15}$ &$7.20^{+1.61}_{-1.34}$\\
{\phantom{\Large{l}}}\raisebox{+.2cm}{\phantom{\Large{j}}}
$\Upsilon(1S)\rightarrow B_s^{\ast0}\overline D^{\ast0}+\overline B_s^{\ast0}D^{\ast0}$&$b_2$&$10^{-14}$ &$2.82^{+0.64}_{-0.53}$\\
\hline\hline
\end{tabular*}
\end{table}

\begin{table}[ht]
 \caption{The branching ratios of non-leptonic decays of $\Upsilon(1S)$ for the Cabibbo-suppressed and color-suppressed (${\rm CaS\otimes CoS}$) $b\rightarrow uq_2\bar q_3$ channels.}
 \setlength{\tabcolsep}{0.5cm}
 \centering
\begin{tabular*}{\textwidth}{@{}@{\extracolsep{\fill}}ccc}
\hline \hline
Channel &Unit&Ours \\ \hline
{\phantom{\Large{l}}}\raisebox{+.2cm}{\phantom{\Large{j}}}
$\Upsilon(1S)\rightarrow B_s^{0}\pi^{0}+\overline B_s^{0}\pi^{0}$  &$10^{-18}$&$6.40^{+1.45}_{-1.25}$\\
{\phantom{\Large{l}}}\raisebox{+.2cm}{\phantom{\Large{j}}}
$\Upsilon(1S)\rightarrow B_s^{0}\rho^{0}+\overline B_s^{0}\rho^{0}$ &$10^{-17}$ &$2.01^{+0.46}_{-0.39}$\\
{\phantom{\Large{l}}}\raisebox{+.2cm}{\phantom{\Large{j}}}
$\Upsilon(1S)\rightarrow B_s^{0}\omega^{0}+\overline B_s^{0}\omega^{0}$  &$10^{-17}$&$1.83^{+0.42}_{-0.35}$\\
{\phantom{\Large{l}}}\raisebox{+.2cm}{\phantom{\Large{j}}}
$\Upsilon(1S)\rightarrow B_s^{0}\eta+\overline B_s^{0}\eta$  &$10^{-18}$&$4.44^{+1.01}_{-0.85}$\\
{\phantom{\Large{l}}}\raisebox{+.2cm}{\phantom{\Large{j}}}
$\Upsilon(1S)\rightarrow B_s^{0}\eta^\prime+\overline B_s^{0}\eta^\prime$ &$10^{-18}$ &$3.55^{+0.80}_{-0.67}$\\
{\phantom{\Large{l}}}\raisebox{+.2cm}{\phantom{\Large{j}}}
$\Upsilon(1S)\rightarrow B_s^{\ast0}\pi^{0}+\overline B_s^{\ast0}\pi^{0}$ &$10^{-17}$ &$4.08^{+0.95}_{-0.79}$\\
{\phantom{\Large{l}}}\raisebox{+.2cm}{\phantom{\Large{j}}}
$\Upsilon(1S)\rightarrow B_s^{\ast0}\rho^{0}+\overline B_s^{\ast0}\rho^{0}$ &$10^{-16}$ &$1.17^{+0.27}_{-0.25}$\\
{\phantom{\Large{l}}}\raisebox{+.2cm}{\phantom{\Large{j}}}
$\Upsilon(1S)\rightarrow B_s^{\ast0}\omega^{0}+\overline B_s^{\ast0}\omega^{0}$ &$10^{-16}$ &$1.07^{+0.25}_{-0.21}$\\
{\phantom{\Large{l}}}\raisebox{+.2cm}{\phantom{\Large{j}}}
$\Upsilon(1S)\rightarrow B_s^{\ast0}\eta+\overline B_s^{\ast0}\eta$&$10^{-17}$  &$2.85^{+0.66}_{-0.55}$\\
{\phantom{\Large{l}}}\raisebox{+.2cm}{\phantom{\Large{j}}}
$\Upsilon(1S)\rightarrow B_s^{\ast0}\eta^\prime+\overline B_s^{\ast0}\eta^\prime$ &$10^{-17}$ &$2.31^{+0.53}_{-0.44}$\\
{\phantom{\Large{l}}}\raisebox{+.2cm}{\phantom{\Large{j}}}
$\Upsilon(1S)\rightarrow B^{0}\overline D^{0}+\overline B^{0}D^{0}$ &$10^{-17}$ &$2.94^{+0.56}_{-0.53}$\\
{\phantom{\Large{l}}}\raisebox{+.2cm}{\phantom{\Large{j}}}
$\Upsilon(1S)\rightarrow B^{\ast0}\overline D^{0}+\overline B^{\ast0}D^{0}$ &$10^{-16}$ &$2.07^{+0.41}_{-0.35}$\\
{\phantom{\Large{l}}}\raisebox{+.2cm}{\phantom{\Large{j}}}
$\Upsilon(1S)\rightarrow B^{0}\overline D^{\ast0}+\overline B^{0}D^{\ast0}$ &$10^{-16}$ &$1.90^{+0.39}_{-0.33}$\\
{\phantom{\Large{l}}}\raisebox{+.2cm}{\phantom{\Large{j}}}
$\Upsilon(1S)\rightarrow B^{\ast0}\overline D^{\ast0}+\overline B^{\ast0}D^{\ast0}$ &$10^{-16}$&$8.23^{+1.68}_{-1.43}$\\
\hline\hline
\end{tabular*}
\end{table}

\begin{table}[ht]
 \caption{The branching ratios of non-leptonic decays of $\Upsilon(1S)$ for the Cabibbo-favored and color-favored (${\rm CaF\otimes CoF}$) $b\rightarrow cq_2\bar q_3$ channels.}
 \setlength{\tabcolsep}{0.01cm}
 \centering
\begin{tabular*}{\textwidth}{@{}@{\extracolsep{\fill}}ccccccc}
\hline \hline
Channel&Unit &Ours & Ref.~\cite{Dhir}&Ref.~\cite{Sha}&Ref.~\cite{Sun02}&Ref.~\cite{Sun03}\\ \hline
{\phantom{\Large{l}}}\raisebox{+.2cm}{\phantom{\Large{j}}}
$\Upsilon(1S)\rightarrow B_{c}^{+}\pi^{-}+B_{c}^{-}\pi^{+}$&$10^{-12}$ &$9.19^{+2.07}_{-1.73}$ & $32^{+2}_{-2}$&66&67.8&\\
{\phantom{\Large{l}}}\raisebox{+.2cm}{\phantom{\Large{j}}}
$\Upsilon(1S)\rightarrow B_{c}^{*+}\pi^{-}+B_{c}^{*-}\pi^{+}$ &$10^{-11}$&$5.15^{+1.14}_{-0.96}$&&& &\\
{\phantom{\Large{l}}}\raisebox{+.2cm}{\phantom{\Large{j}}}
$\Upsilon(1S)\rightarrow B_{c}^{+}\rho^{-}+B_{c}^{-}\rho^{+}$&$10^{-11}$&$2.29^{+0.52}_{-0.43}$ &$13.0^{+0.6}_{-0.6}$&17.6&20.8&\\
{\phantom{\Large{l}}}\raisebox{+.2cm}{\phantom{\Large{j}}}
$\Upsilon(1S)\rightarrow B_{c}^{*+}\rho^{-}+B_{c}^{*-}\rho^{+}$&$10^{-10}$ &$1.27^{+0.28}_{-0.24}$  &&&&\\
{\phantom{\Large{l}}}\raisebox{+.2cm}{\phantom{\Large{j}}}
$\Upsilon(1S)\rightarrow B_{c}^{+}D_s^{-}+B_{c}^{-}D_s^{+}$ &$10^{-11}$ &$3.97^{+0.71}_{-0.62}$&$9.4^{+0.2}_{-0.2}$ &15.2&&54.2\\
{\phantom{\Large{l}}}\raisebox{+.2cm}{\phantom{\Large{j}}}
$\Upsilon(1S)\rightarrow B_{c}^{+}D_s^{*-}+B_{c}^{-}D_s^{*+}$&$10^{-10}$ & $2.78^{+0.50}_{-0.43}$&$3.58^{+0.18}_{-0.08}$&5.14&&\\
{\phantom{\Large{l}}}\raisebox{+.2cm}{\phantom{\Large{j}}}
$\Upsilon(1S)\rightarrow B_{c}^{*+}D_s^{-}+B_{c}^{*-}D_s^{+}$ &$10^{-10}$& $2.86^{+0.51}_{-0.45}$&&&&\\
{\phantom{\Large{l}}}\raisebox{+.2cm}{\phantom{\Large{j}}}
$\Upsilon(1S)\rightarrow B_{c}^{*+}D_s^{*-}+B_{c}^{*-}D_s^{*+}$ &$10^{-10}$ &$8.54^{+1.48}_{-1.28}$ &&&&\\
\hline\hline
\end{tabular*}
\end{table}


\begin{table}[ht]
 \caption{The branching ratios of non-leptonic decays of $\Upsilon(1S)$ for the Cabibbo-suppressed and color-favored (${\rm CaS\otimes CoF}$, labeled by $b_3$) or Cabibbo-favored and color-suppressed (${\rm CaF\otimes CoS}$, labeled by $b_4$) $b\rightarrow cq_2\bar q_3$ channels.}
 \setlength{\tabcolsep}{0cm}
\centering
\begin{tabular*}{\textwidth}{@{}@{\extracolsep{\fill}}cccccccc}
\hline \hline
Channel &Type&Unit&Ours & Ref.~\cite{Dhir}&Ref.~\cite{Sha}&Ref.~\cite{Sun02}&Ref.~\cite{Sun03}\\ \hline
{\phantom{\Large{l}}}\raisebox{+.2cm}{\phantom{\Large{j}}}
$\Upsilon(1S)\rightarrow B_{c}^{+}K^{-}+B_{c}^{-}K^{+}$& $b_3$& $10^{-13}$&$7.13^{+1.59}_{-1.33}$&$26$& $48$&50.2&\\
{\phantom{\Large{l}}}\raisebox{+.2cm}{\phantom{\Large{j}}}
$\Upsilon(1S)\rightarrow B_{c}^{+}K^{*-}+B_{c}^{-}K^{*+}$&$b_3$&$10^{-12}$ &$1.82^{+0.40}_{-0.33}$ & $7.0^{+4.0}_{-4.0}$&10&10.5&\\
{\phantom{\Large{l}}}\raisebox{+.2cm}{\phantom{\Large{j}}}
$\Upsilon(1S)\rightarrow B_{c}^{*+}K^{-}+B_{c}^{*-}K^{+}$&$b_3$& $10^{-12}$&$4.04^{+0.89}_{-0.74}$ &&&&\\
{\phantom{\Large{l}}}\raisebox{+.2cm}{\phantom{\Large{j}}}
$\Upsilon(1S)\rightarrow B_{c}^{*+}K^{*-}+B_{c}^{*-}K^{*+}$&$b_3$&$ 10^{-12}$ &$8.75^{+1.88}_{-1.57}$ &&&&\\
{\phantom{\Large{l}}}\raisebox{+.2cm}{\phantom{\Large{j}}}
$\Upsilon(1S)\rightarrow B_{c}^{+}D^{-}+B_{c}^{-}D^{+}$&$b_3$&$10^{-12}$ & $1.32^{+0.24}_{-0.21}$&$3^{+2}_{-0}$&6.2&&19.6\\
{\phantom{\Large{l}}}\raisebox{+.2cm}{\phantom{\Large{j}}}
$\Upsilon(1S)\rightarrow B_{c}^{+}D^{\ast-}+B_{c}^{-}D^{\ast+}$&$b_3$&$10^{-11}$& $1.09^{+0.20}_{-0.17}$ &$1.46^{+0.06}_{-0.04}$&2.2&&\\
{\phantom{\Large{l}}}\raisebox{+.2cm}{\phantom{\Large{j}}}
$\Upsilon(1S)\rightarrow B_{c}^{*+}D^{-}+B_{c}^{*-}D^{+}$&$b_3$ &$10^{-12}$&$9.16^{+1.69}_{-1.45}$&& &&\\
{\phantom{\Large{l}}}\raisebox{+.2cm}{\phantom{\Large{j}}}
$\Upsilon(1S)\rightarrow B_{c}^{*+}D^{\ast-}+B_{c}^{*-}D^{\ast+}$&$b_3$&$10^{-11}$ &$3.48^{+0.62}_{-0.54}$ &&&&\\
{\phantom{\Large{l}}}\raisebox{+.2cm}{\phantom{\Large{j}}}
$\Upsilon(1S)\rightarrow B_{s}^{0}~\eta_c+\overline B_{s}^{0}~\eta_c$&$b_4$&$10^{-13}$& $7.82^{+1.55}_{-1.33}$ &&$340$&&\\
{\phantom{\Large{l}}}\raisebox{+.2cm}{\phantom{\Large{j}}}
$\Upsilon(1S)\rightarrow B_{s}^{\ast0}~\eta_c+\overline B_{s}^{\ast0}~\eta_c$ &$b_4$&$10^{-12}$&$6.85^{+1.40}_{-1.18}$ &&&&\\
{\phantom{\Large{l}}}\raisebox{+.2cm}{\phantom{\Large{j}}}
$\Upsilon(1S)\rightarrow B_{s}^{0}J/\psi+\overline B_{s}^{0}J/\psi$&$b_4$& $10^{-12}$&$4.64^{+0.97}_{-0.82}$ &&$186$&&\\
{\phantom{\Large{l}}}\raisebox{+.2cm}{\phantom{\Large{j}}}
$\Upsilon(1S)\rightarrow B_{s}^{\ast0}J/\psi+\overline B_{s}^{\ast0}J/\psi$&$b_4$&$10^{-11} $&$1.43^{+0.29}_{-0.24}$ &&&&\\
{\phantom{\Large{l}}}\raisebox{+.2cm}{\phantom{\Large{j}}}
$\Upsilon(1S)\rightarrow B^0D^0+\overline B^0\overline D^0$&$b_4$&1$0^{-14}$ & $5.45^{+0.69}_{-0.69}$&&$2000$&&\\
{\phantom{\Large{l}}}\raisebox{+.2cm}{\phantom{\Large{j}}}
$\Upsilon(1S)\rightarrow B^0D^{\ast0}+\overline B^0\overline D^{\ast0}$&$b_4$&$ 10^{-13}$&$3.52^{+0.50}_{-0.45}$ &&&&\\
{\phantom{\Large{l}}}\raisebox{+.2cm}{\phantom{\Large{j}}}
$\Upsilon(1S)\rightarrow B^{\ast0}D^0+\overline B^{\ast0}\overline D^0$&$b_4$& $10^{-13}$& $3.84^{+0.76}_{-0.66}$&&&&\\
{\phantom{\Large{l}}}\raisebox{+.2cm}{\phantom{\Large{j}}}
$\Upsilon(1S)\rightarrow B^{\ast0}D^{\ast0}+\overline B^{\ast0}\overline D^{\ast0}$& $b_4$&$10^{-12}$&$1.53^{+0.31}_{-0.27}$ &&&&\\
\hline\hline
\end{tabular*}
\end{table}

\begin{table}[ht]
 \caption{The branching ratios of non-leptonic decays of $\Upsilon(1S)$ for the Cabibbo-suppressed and color-suppressed (${\rm CaS\otimes CoS}$) $b\rightarrow cq_2\bar q_3$ channels.}
 \setlength{\tabcolsep}{0.1cm}
 \centering
\begin{tabular*}{\textwidth}{@{}@{\extracolsep{\fill}}cccc}
\hline \hline
Channel &Unit&Ours &Ref.~\cite{Sha} \\ \hline
{\phantom{\Large{l}}}\raisebox{+.2cm}{\phantom{\Large{j}}}
$\Upsilon(1S)\rightarrow B_s^0 D^0+\overline B_s^0\overline D^0$ &$10^{-15}$&$5.38^{+1.18}_{-0.99}$&$500$ \\
{\phantom{\Large{l}}}\raisebox{+.2cm}{\phantom{\Large{j}}}
$\Upsilon(1S)\rightarrow B_s^{0} D^{\ast0}+\overline B_s^0\overline D^{\ast0}$& $10^{-14}$&$3.48^{+0.80}_{-0.66}$ &\\
{\phantom{\Large{l}}}\raisebox{+.2cm}{\phantom{\Large{j}}}
$\Upsilon(1S)\rightarrow B_s^{\ast0} D^0+\overline B_s^{\ast0}\overline D^0$&$10^{-14}$ & $3.72^{+0.83}_{-0.69}$&\\
{\phantom{\Large{l}}}\raisebox{+.2cm}{\phantom{\Large{j}}}
$\Upsilon(1S)\rightarrow B_s^{\ast0} D^{\ast0}+\overline B_s^{\ast0}\overline D^{\ast0}$ &$10^{-13}$& $1.46^{+0.33}_{-0.28}$& \\
{\phantom{\Large{l}}}\raisebox{+.2cm}{\phantom{\Large{j}}}
$\Upsilon(1S)\rightarrow B^0~\eta_c+\overline B^0~\eta_c$ &$10^{-14}$&$2.52^{+0.47}_{-0.41}$&$200$ \\
{\phantom{\Large{l}}}\raisebox{+.2cm}{\phantom{\Large{j}}}
$\Upsilon(1S)\rightarrow B^0 J/\psi+\overline B^0 J/\psi$&$10^{-13}$ &$1.43^{+0.29}_{-0.24}$& $106$\\
{\phantom{\Large{l}}}\raisebox{+.2cm}{\phantom{\Large{j}}}
$\Upsilon(1S)\rightarrow B^{\ast0}~\eta_c+\overline B^{\ast0}~\eta_c$& $10^{-13}$&$2.16^{+0.42}_{-0.36}$& \\
{\phantom{\Large{l}}}\raisebox{+.2cm}{\phantom{\Large{j}}}
$\Upsilon(1S)\rightarrow B^{\ast0} J/\psi+\overline B^{\ast0} J/\psi$ &$10^{-13}$&$4.62^{+0.92}_{-0.78}$& \\
\hline\hline
\end{tabular*}
\end{table}

\section{Acknowledgments}

This paper was supported in part by the National Natural Science
Foundation of China (NSFC) under Grant No.~11405037, No.~11505039 and No.~11575048, and in part by PIRS of HIT No.~A201409 and No.~B201506.

 \begin{appendices}
 \section*{Appendix A. The Explicit Form of $\varphi^{++}$}
The positive part of the wave function for the $1^-$ state has the form
\begin{equation}
\begin{aligned}
\varphi^{++}_{1^-}(q_\perp)&=(q_\perp\cdot\epsilon)\left[A_1(q_\perp)+\frac{\slashed{P}}{M}A_2(q_\perp)
+\frac{\slashed{q}_\perp}{M}A_3(q_\perp)+\frac{\slashed{P}\slashed{q}_\perp}{M^2}A_4(q_\perp)\right]\\
&~~~~~+ M\slashed\epsilon\left[A_5(q_\perp)+\frac{\slashed{P}}{M}A_6(q_\perp)
+\frac{\slashed{q}_\perp}{M}A_7(q_\perp)+\frac{\slashed{P}\slashed{q}_\perp}{M^2}A_8(q_\perp)\right],
\end{aligned}
\end{equation}
where $A_i$s are defined as
\begin{equation}
\begin{aligned}
&A_{1}=\frac{(\omega_{1}+\omega_{2})q_{\perp}^{2}f_{3}+(m_{1}+m_{2})q_{\perp}^{2}f_{4}+2M^{2}\omega_{2}f_{5}-2M^{2}m_{2}f_{6}}{2M(m_{1}\omega_{2}+m_{2}\omega_{1})},\\
&A_{2}=\frac{(m_{1}-m_{2})q_{\perp}^{2}f_{3}+(\omega_{1}-\omega_{2})q_{\perp}^{2}f_{4}-2M^{2}m_{2}f_{5}+2M^{2}\omega_{2}f_{6}}{2M(m_{1}\omega_{2}+m_{2}\omega_{1})},\\
&A_{3}=\frac{1}{2}(f_{3}+\frac{m_{1}+m_{2}}{\omega_{1}+\omega_{2}}f_{4}-\frac{2M^{2}}{m_{1}\omega_{2}+m_{2}\omega_{1}}f_{6}),\\
&A_{4}=\frac{1}{2}(\frac{\omega_{1}+\omega_{2}}{m_{1}+m_{2}}f_{3}+f_{4}-\frac{2M^{2}}{m_{1}\omega_{2}+m_{2}\omega_{1}}f_{5}),\\
&A_{5}=\frac{1}{2}(f_{5}-\frac{\omega_{1}+\omega_{2}}{m_{1}+m_{2}}f_{6}),~~~~~A_{6}=\frac{1}{2}(-\frac{m_{1}+m_{2}}{\omega_{1}+\omega_{2}}f_{5}+f_{6}),\\
&A_{7}=A_{5}\frac{M(\omega_{1}-\omega_{2})}{m_{1}\omega_{2}+m_{2}\omega_{1}},~~~~~~~~A_{8}=A_{6}\frac{M(\omega_{1}+\omega_{2})}{m_{1}\omega_{2}+m_{2}\omega_{1}}.\\
\end{aligned}
\end{equation}
And for the $0^-$ final state, the wave function has the form
\begin{equation}
\begin{aligned}
\varphi^{++}_{0^-}(q_{f\perp})&=\left[B_1(q_{f\perp})+\frac{\slashed{P}_f}{M_f}B_2(q_{f\perp})
+\frac{\slashed{q}_{f\perp}}{M_f}B_3(q_{f\perp})+\frac{\slashed{P}_f\slashed{q}_{f\perp}}{M_f^2}B_4(q_{f\perp})\right]\gamma_5,
\end{aligned}
\end{equation}
where
\begin{equation}
\begin{aligned}
&B_{1}=\frac{M_{f}}{2}(\frac{\omega^\prime_{1}+\omega^\prime_{2}}{m^\prime_{1}+m^\prime_{2}}g_{1}+g_{2}),\\
&B_{2}=\frac{M_{f}}{2}(g_{1}+\frac{m^\prime_{1}+m^\prime_{2}}{\omega^\prime_{1}+\omega^\prime_{2}}g_{2}),\\
&B_{3}=-\frac{M_{f}(\omega^\prime_{1}-\omega^\prime_{2})}{m^\prime_{1}\omega^\prime_{2}+m^\prime_{2}\omega^\prime_{1}}B_{1},\\
&B_{4}=-\frac{M_{f}(\omega^\prime_{1}+\omega^\prime_{2})}{m^\prime_{1}\omega^\prime_{2}+m^\prime_{2}\omega^\prime_{1}}B_{2}.\\
\end{aligned}
\end{equation}
In the above equations, we have used the definitions: $q^\mu_\perp=q^\mu-\frac{P\cdot q}{M^2}P^\mu$, $q^\mu_{f\perp}=q_f^\mu-\frac{P\cdot q_f}{M^2}P^\mu$, $\omega_i=\sqrt{m_i^2-q_\perp^2}$, and $\omega^\prime_i=\sqrt{m^{\prime2}_i-q_{f\perp}^2}$.  $f_i$s and $g_i$s are functions of $q_\perp^2$ and $q_{f\perp}^2$, respectively.

\section*{Appendix B. The Explicit Forms of $\alpha$, $\beta_{\pm\pm}$, and $\gamma$}
 $\alpha$, $\beta_{\pm\pm}$, and $\gamma$ are expressed by the form factors. For the $1^{--}\rightarrow 0^-$ process, they have the forms
\begin{equation}
\begin{aligned}
\alpha = \frac{4M^2\vec P_f^2}{(M+M_f)^2} s_1^2 + (M+M_f)^2 s_2^2,
\end{aligned}
\end{equation}
\begin{equation}
\begin{aligned}
\beta_{++} =&\frac{\vec P_f^2}{4M^2}(s_3-s_4)^2 + \frac{1}{(M+M_f)^2}(2ME_f-M^2-M_f^2)s_1^2+ \frac{1}{2M^2}(M-E_f)\\&\times(M+M_f)s_2(s_3-s_4) + \frac{1}{4M^2}(M+M_f)^2s_2^2,
\end{aligned}
\end{equation}
\begin{equation}
\begin{aligned}
\beta_{+-} = \beta_{-+}& =-\frac{M_f^2}{4M^2} s_3^2 -\frac{\vec P_f^2}{4M^2} s_4^2 + (\frac{E_f}{2M}s_3 - \frac{M+M_f}{2M}s_2)^2 + \frac{M-M_f}{M+M_f} s_1^2 \\
&+ \frac{M+M_f}{2M} s_2s_4,
\end{aligned}
\end{equation}
\begin{equation}
\begin{aligned}
\beta_{--} =&\frac{\vec P_f^2}{4M^2}(s_3+s_4)^2 - \frac{1}{(M+M_f)^2}(2ME_f+M^2+M_f^2)s_1^2- \frac{1}{2M^2}(M+E_f)\\&\times(M+M_f)s_2(s_3+s_4) + \frac{1}{4M^2}(M+M_f)^2s_2^2,
\end{aligned}
\end{equation}
\begin{equation}
\begin{aligned}
\gamma = -2s_1s_2.
\end{aligned}
\end{equation}

For the $1^{--}\rightarrow 1^-$ process, they have forms
\begin{equation}
\begin{aligned}
\alpha =& -\frac{2E_f\vec P_f^2}{MM_f^2} (M^2 h_3h_5 + M_f^2 h_4h_6) + 2P_f^2 (h_3h_6 + h_4h_5) + (M_f^2 + E_f^2) (\frac{M^2}{M_f^2} h_5^2 \\
&+ h_6^2) + \vec P_f^2 (\frac{M^2}{M_f^2} t_3^2 + t_4^2) + \vec P_f^4 (\frac{h_3^2}{M_f^2} + \frac{h_4^2}{M^2}) - 4ME_f h_5 h_6,
\end{aligned}
\end{equation}

\begin{equation}
\begin{aligned}
\beta_{++} &= \frac{\vec P_f^2}{2M^2} (h_1 - h_2)^2 + \frac{\vec P_f^4}{4M^2M_f^2} (t_1 + t_2)^2 + \frac{E_f \vec P_f^2}{2MM_f^2} (t_1+t_2)(t_5+t_6) + (\frac{E_f^2}{4M_f^2}\\
&+\frac{1}{2})(t_5+t_6)^2 + \frac{1}{2M^2}(2ME_f-M^2 - M_f^2)(h_4h_5+h_3h_6) + \frac{\vec P_f^2}{4M^4M_f^2} (2ME_f\\
&-M_f^2 - M^2) (M^2 h_3^2 + M_f^2 h_4^2) + \frac{E_f}{2M^3M_f^2}(M^2+ M_f^2 -2ME_f)(M^2h_3h_5\\
& + M_f^2h_4h_6)+(1-\frac{E_f}{M})(h_1-h_2)h_5 + \frac{1}{M^2}(M_f^2 - ME_f)(h_1-h_2)h_6 + \frac{\vec P_f^2}{4M_f^2}\\
&\times(t_3^2+t_4^2) - h_5h_6 + \frac{\vec P_f^2}{2M_f^2}(1-\frac{E_f}{M})(t_1+t_2)t_3 + \frac{\vec P_f^2}{2M^2M_f^2}(M_f^2 - ME_f)\\
&\times(t_1+t_2)t_4 +\frac{E_f}{2MM_f^2}(M_f^2-ME_f)(t_5+t_6)t_4 + \frac{ME_f}{2M_f^2}(1-\frac{E_f}{M})(t_5+t_6)t_3\\
&+\frac{1}{2M_f^2}(M_f^2-ME_f)(1-\frac{E_f}{M})t_3t_4 + \frac{1}{4M_f^2}(-M^2+2ME_f+M_f^2)h_5^2 + \frac{1}{4M^2}\\
&\times(-M_f^2+2ME_f+M^2)h_6^2,
\end{aligned}
\end{equation}
\begin{equation}
\begin{aligned}
\beta_{+-}&=\beta_{-+}=\frac{E_f\vec P_f^2}{2MM_f^2}(-t_1t_3+t_1t_5+t_2t_4-t_2t_6) +\frac{\vec P_f^2}{2M^2}(h_1^2-h_2^2) + \frac{\vec P_f^2}{4M^4M_f^2}\\
&\times(M^2-M_f^2)(M^2h_3^2 + M_f^2h_4^2) + \frac{\vec P_f^2}{2M^2M_f^2}(M_f^2t_1t_4 - M^2t_2t_3) + \frac{\vec P_f^2}{4M_f^2}(t_3^2-t_4^2) \\
&- \frac{E_f^2}{2M_f^2}(t_3t_5 - t_4t_6) + (\frac{E_f^2}{4M_f^2}+\frac{1}{2})(t_5^2-t_6^2) + \frac{\vec P_f^4}{4M^2M_f^2}(t_1^2-t_2^2) - \frac{E_f}{M}(h_1h_5\\
&+h_2h_6) +\frac{E_f}{2M^3M_f^2}(M_f^2-M^2)(M^2h_3h_5+M_f^2h_4h_6-M^2t_3t_4)-\frac{E_f}{2MM_f^2}\\
&\times(M^2t_3t_6-M_f^2t_4t_5) +\frac{1}{M^2}(M_f^2h_1h_6+M^2h_2h_5) + \frac{1}{2M^2}(M^2-M_f^2)(h_3h_6\\
&+h_4h_5)+ \frac{1}{4M^2M_f^2}(M^2+M_f^2)(M^2h_5^2-M_f^2h_6^2),
\end{aligned}
\end{equation}
\begin{equation}
\begin{aligned}
\beta_{--} &= \frac{\vec P_f^2}{2M^2} (h_1 + h_2)^2 + \frac{\vec P_f^4}{4M^2M_f^2} (t_1 - t_2)^2 + \frac{E_f \vec P_f^2}{2MM_f^2} (t_1-t_2)(t_5-t_6) + (\frac{E_f^2}{4M_f^2}\\
&+\frac{1}{2})(t_5-t_6)^2 - \frac{1}{2M^2}(2ME_f+M^2 + M_f^2)(h_4h_5+h_3h_6) - \frac{\vec P_f^2}{4M^4M_f^2} (2ME_f\\
&+M_f^2 + M^2) (M^2 h_3^2 + M_f^2 h_4^2) + \frac{E_f}{2M^3M_f^2}(M^2+ M_f^2 +2ME_f)(M^2h_3h_5\\
& + M_f^2h_4h_6)-(1+\frac{E_f}{M})(h_1+h_2)h_5 + \frac{1}{M^2}(M_f^2 + ME_f)(h_1+h_2)h_6 + \frac{\vec P_f^2}{4M_f^2}\\
&\times(t_3^2+t_4^2) + h_5h_6 - \frac{\vec P_f^2}{2M_f^2}(1+\frac{E_f}{M})(t_1-t_2)t_3 + \frac{\vec P_f^2}{2M^2M_f^2}(M_f^2 + ME_f)\\
&\times(t_1-t_2)t_4 +\frac{E_f}{2MM_f^2}(M_f^2+ME_f)(t_5-t_6)t_4 - \frac{ME_f}{2M_f^2}(1+\frac{E_f}{M})(t_5-t_6)t_3\\
&-\frac{1}{2M_f^2}(M_f^2+ME_f)(1+\frac{E_f}{M})t_3t_4 + \frac{1}{4M_f^2}(-M^2-2ME_f+M_f^2)h_5^2 + \frac{1}{4M^2}\\
&\times(-M_f^2-2ME_f+M^2)h_6^2,
\end{aligned}
\end{equation}
\begin{equation}
\begin{aligned}
\gamma = \frac{\vec P_f^2}{M_f^2}(h_3t_3-h_4t_4)-\frac{E_fM}{M_f^2}h_5t_3+\frac{E_f}{M}h_6t_4 -h_5t_4+h_6t_3.
\end{aligned}
\end{equation}

\end{appendices}

\end{document}